\documentclass[lettersize,journal]{IEEEtran}
\DeclareMathSizes{10}{10}{6}{5}
\usepackage{amsmath,amsfonts}
\usepackage{algorithmic}
\usepackage{algorithm} 
\usepackage{array}
\usepackage[caption=false,font=normalsize,labelfont=sf,textfont=sf]{subfig}
\usepackage{textcomp}
\usepackage{stfloats}
\usepackage{url}
\usepackage{verbatim}
\usepackage{graphicx}
\usepackage{cite}
\usepackage{CJKutf8}
\usepackage{mathrsfs,amsthm,amsmath,amssymb,lipsum}
\usepackage{cuted,mwe,float,caption}
\usepackage{stfloats}
\usepackage{verbatim}
\usepackage{multirow}
\usepackage{float}
\usepackage{subcaption}
\newcommand{\Rmnum}[1]{\uppercase\expandafter{\romannumeral #1}}

\begin{document}
\begin{CJK}{UTF8}{gbsn}

\title{Heterogeneous Secure Transmissions in IRS-Assisted NOMA Communications: CO-GNN Approach}

\author{Linlin Liang, ~\IEEEmembership{Member,~IEEE}, Zongkai Tian, Haiyan Huang,~\IEEEmembership{Member,~IEEE}, Xiaoyan Li, Zhisheng Yin, ~\IEEEmembership{Member,~IEEE},  Dehua Zhang, Nina Zhang, Wenchao Zhai

\thanks{Linlin Liang, Zongkai Tian, Xiaoyan Li are with the School of Cyber Engineering and Zhisheng Yin is with the School of Telecommunications Engineering, Xidian University, Xi’an, 710071, China (email: llliang@xidian.edu.cn;zktian@stu.xidian.edu.cn; xiaoyanleece@stu.xidian.edu.cn; zsyin@xidian.edu.cn).

Haiyan Huang is with the School of Electronic and Information Engineering, Lanzhou Jiaotong University, Lanzhou, 730070, China (email: huanghaiyan@mail.lzjtu.cn).

Wenchao Zhai is with the College of Information Engineering, China Jiliang University, Hangzhou, 310018, China (email:zhaiwenchao@cjlu.edu.cn).

Nina Zhang is with the Comprehensive Information Support Center, Shaanxi General Staff of PAP, Xi’an, 710054, China (email: zhangnina0301@yeah.net).

Dehua Zhang is with the School of Artificial Intelligence, Henan University, Zhengzhou, 450046, China (email: dhuazhang@vip.henu.edu.cn).}% <-this % stops a space
\thanks{Manuscript received September xx, 2024. This work was supported by the National Natural Science Foundation of China under Grants 62001359 and 62461032, National basic scientific research of China under Grants JCKY2023110C099, and by the Key Science and Technology Research Project of Henan Province under Grants 232102211059. (\textit{Corresponding author:} Zhisheng Yin and Haiyan Huang.)}}

\markboth{Journal of \LaTeX\ Class Files,~Vol.~14, No.~8, August~2024}%
{Shell \MakeLowercase{\textit{et al.}}: A Sample Article Using IEEEtran.cls for IEEE Journals}


\maketitle

\begin{abstract}
Intelligent Reflecting Surfaces (IRS) enhance spectral efficiency by adjusting reflection phase shifts, while Non-Orthogonal Multiple Access (NOMA) increases system capacity. Consequently, IRS-assisted NOMA communications have garnered significant research interest. However, the passive nature of the IRS, lacking authentication and security protocols, makes these systems vulnerable to external eavesdropping due to the openness of electromagnetic signal propagation and reflection. NOMA's inherent multi-user signal superposition also introduces internal eavesdropping risks during user pairing. This paper investigates secure transmissions in IRS-assisted NOMA systems with heterogeneous resource configuration in wireless networks to mitigate both external and internal eavesdropping. To maximize the sum secrecy rate of legitimate users, we propose a combinatorial optimization graph neural network (CO-GNN) approach to jointly optimize beamforming at the base station, power allocation of NOMA users, and phase shifts of IRS for dynamic heterogeneous resource allocation, thereby enabling the design of dual-link or multi-link secure transmissions in the presence of eavesdroppers on the same or heterogeneous links. The CO-GNN algorithm simplifies the complex mathematical problem-solving process, eliminates the need for channel estimation, and enhances scalability. Simulation results demonstrate that the proposed algorithm significantly enhances the secure transmission performance of the system.  
\end{abstract}


\section{Introduction}

\IEEEPARstart{T}{he} escalating demand for high data capacity and low transmission latency in wireless communication has led to significant proliferation in recent decades. To address the challenges posed by the frequent and substantial information exchange, non-orthogonal multiple access (NOMA) has garnered widespread attention in both academia and industry due to its superior spectrum efficiency, extensive connectivity, and enhanced user fairness. While NOMA shows promise in strengthening spectral efficiency and user fairness, its vulnerability to passive attacks such as eavesdropping due to the open nature of wireless communication poses a critical challenge in ensuring the security of NOMA transmissions.

Physical layer security (PLS) techniques are specifically designed to enhance the confidentiality and integrity of transmitted information by leveraging the physical attributes of wireless communication channels. In this regard, the intelligent reflecting surface (IRS) has garnered significant attention for its ability to improve the transmission performance and security of NOMA networks. The individual elements on the IRS can manipulate the phase of the incident signal independently, thereby enhancing the received signal from the base station to users and thwarting potential eavesdroppers from intercepting information. As a result of the advantages provided by IRS technology, IRS-enhanced wireless networks are being explored as a viable solution to enhance system transmission and security.

\subsection{Related Works}
%IRS-NOMA 引入
%\textit{et al.}\cite{7514758}
Existing studies have conducted in-depth exploration of performance analysis and resource configuration in IRS-assited-NOMA systems. Sun \textit{et al.} \cite{9566699} first compared the performance differences between ideal IRS (with continuous phase shifts) and non-ideal IRS (with discrete phase shifts) in NOMA networks, deriving closed-form expressions for outage probability and revealing the critical impact of IRS phase precision on system capacity. Subsequently, researchers employed methods such as alternating optimization (AO) and penalty dual decomposition (PDD) \cite{9240028,10149169,9139273} to jointly optimize BS beamforming and IRS reflection matrices to maximize the system sum rate. Furthermore, some studies utilized a two-stage approach to optimize NOMA power allocation, IRS phase shifts, and other parameters by transforming non-convex problems into convex ones for analysis \cite{10086618,9531372}. For example, Khan \textit{et al.} \cite{10086618} decoupled the original optimization problem into two subproblems: power allocation and passive beamforming, solved via the inner approximation method and convex optimization, respectively. However, existing works predominantly focus on single objectives and fail to jointly optimize heterogeneous resources, relying on traditional convex optimization frameworks that struggle to meet real-time requirements in dynamic eavesdropping scenarios.

With advancements in deep learning technologies, neural networks have gradually been applied to parameter optimization in IRS-NOMA systems. Chandan \textit{et al.} \cite{10236498} constructed a deep neural network (DNN) to predict outage probability and ergodic rates under hardware impairments while considering residual hardware damage. Ridho \textit{et al.} \cite{10129167} designed a deep learning model to jointly optimize precoding matrices and IRS phase shifts, but their user pairing strategy remains confined to static scenarios. Gao \textit{et al.} \cite{9594701} employed long short-term memory, a K-means-based Gaussian mixture model, and deep Q-networks to maximize the sum rate of all users. In reinforcement learning, Chen \textit{et al.} \cite{9880822} proposed a multi-agent deep reinforcement learning framework to optimize IRS phases and downlink power through information interaction, aiming to maximize overall energy efficiency. Yu \textit{et al.} \cite{10279340,9968198} introduced a Lyapunov-function-based mixed-integer deep deterministic policy gradient algorithm within a multi-agent reinforcement learning framework to enhance communication spectral efficiency. However, these studies do not explore dynamic resource allocation mechanisms tailored for secrecy rate maximization. Although some works combine traditional optimization algorithms, such as AO \cite{9896673}, to reduce complexity, their generalization capabilities remain limited by fixed system models, making them unsuitable for security requirements in heterogeneous links.

To address eavesdropping threats in IRS-NOMA systems, physical layer security techniques have become a research hotspot. Some studies counteract potential eavesdropping through coordinated jamming or artificial noise injection \cite{9520776,9627968,9266086,9739715}. Wang \textit{et al.} \cite{9520776} proposed an IRS-assisted artificial noise scheme to suppress passive eavesdropping by jointly transmitting NOMA signals and jamming signals. Several researchers adopted secrecy outage probability (SOP) as a security metric to analyze system security \cite{9606864,9619950}. Lu \textit{et al.} \cite{9524501} optimized transmit power and IRS reflect beamforming with SOP as the objective. Additionally, Zhang \textit{et al.} \cite{10278724} decomposed active and passive beamforming problems using AO to maximize the secrecy rate of primary users but neglected the collaborative attack risks posed by internal and external eavesdroppers. Han \textit{et al.} \cite{9896889} considered scenarios with both internal and external eavesdroppers, jointly optimizing active and passive beamforming to maximize the secrecy rate while minimizing legitimate users’ transmit power and increasing artificial noise power to disrupt eavesdroppers. Furthermore, Guo \textit{et al.} \cite{10376206} integrated unmanned aerial vehicles (UAVs) with IRS and adopted a double deep Q-network algorithm to learn online UAV trajectory design strategies for maximizing the system secrecy rate. Despite leveraging traditional optimization methods to enhance security performance, existing works heavily depend on precise channel state information (CSI) and suffer from high computational complexity, hindering their applicability in large-scale dynamic network environments.

\begin{table*}[!ht]
    \centering
    \caption{Comparisons of proposed and exiting works}
    \setlength{\tabcolsep}{0.85mm}{
    \begin{tabular}{|l|l|l|l|l|l|l|l|l|l|l|l|l|l|l|l|l|l|l|l|l|l|l|l|l|l|l|l|}
    \hline
        CTX & [1] & [2] & [3] & [4] & [5] & [6] & [7] & [8] & [9] & [10] & [11] & [12] & [13] & [14] & [15] & [16] & [17] & [18] & [19] & [20] & [21] & [22] & [23] & [24] & [25] & [26] & [27] \\ \hline
        $\mathcal{A}$ & \checkmark & \checkmark & \checkmark & \checkmark & \checkmark & \checkmark & \checkmark & \checkmark & \checkmark & \checkmark & \checkmark & \checkmark & \checkmark & \checkmark & \checkmark & \checkmark & \checkmark & \checkmark & \checkmark & \checkmark & \checkmark & \checkmark & \checkmark & \checkmark & \checkmark & \checkmark & \checkmark \\ \hline
        $\mathcal{B}$ & \checkmark & \checkmark & \checkmark & \checkmark & \checkmark & \checkmark & \checkmark & \checkmark & \checkmark & \checkmark & \checkmark & \checkmark & \checkmark & \checkmark & \checkmark & \checkmark & \checkmark & \checkmark & \checkmark & \checkmark & \checkmark & \checkmark & \checkmark & \checkmark & \checkmark & \checkmark & \checkmark \\ \hline
        $\mathcal{C}$ & ~ & \checkmark & \checkmark & \checkmark & \checkmark & \checkmark & \checkmark & \checkmark & ~ & ~ & \checkmark & ~ & ~ & ~ & \checkmark & ~ & \checkmark & \checkmark & \checkmark & \checkmark & \checkmark & ~ & \checkmark & \checkmark & ~ & ~ & ~ \\ \hline
        $\mathcal{D}$ & ~ & \checkmark & \checkmark & \checkmark & ~ & \checkmark & ~ & \checkmark & ~ & \checkmark & ~ & \checkmark & ~ & ~ & ~ & \checkmark & \checkmark & \checkmark & \checkmark & \checkmark & \checkmark & ~ & ~ & \checkmark & \checkmark & \checkmark & ~ \\ \hline
        $\mathcal{E}$ & ~ & ~ & ~ & ~ & \checkmark & \checkmark & \checkmark & ~ & ~ & ~ & \checkmark & \checkmark & ~ & ~ & ~ & \checkmark & ~ & ~ & ~ & ~ & ~ & ~ & ~ & ~ & ~ & \checkmark & \checkmark \\ \hline
        $\mathcal{F}$ & ~ & ~ & ~ & ~ & ~ & ~ & ~ & ~ & ~ & ~ & ~ & ~ & ~ & ~ & ~ & ~ & \checkmark & \checkmark & \checkmark & \checkmark & \checkmark & \checkmark & \checkmark & \checkmark & \checkmark & \checkmark & \checkmark \\ \hline
        $\mathcal{G}$ & ~ & ~ & ~ & ~ & ~ & ~ & ~ & ~ & ~ & ~ & ~ & ~ & ~ & ~ & ~ & ~ & ~ & ~ & ~ & ~ & ~ & \checkmark & ~ & \checkmark & ~ & ~ & ~ \\ \hline
        $\mathcal{H}$ & ~ & ~ & ~ & ~ & ~ & ~ & ~ & ~ & \checkmark & \checkmark & \checkmark & \checkmark & \checkmark & \checkmark & \checkmark & \checkmark & ~ & ~ & ~ & ~ & ~ & ~ & ~ & ~ & ~ & ~ & \checkmark \\ \hline
        % $\mathcal{I}$ & ~ & ~ & ~ & ~ & ~ & ~ & ~ & ~ & \checkmark & ~ & ~ & ~ & ~ & ~ & ~ & ~ & ~ & ~ & ~ & ~ & ~ & ~ & ~ & ~ & ~ & ~ & ~ \\ \hline
        $\mathcal{I}$ & ~ & ~ & ~ & ~ & ~ & ~ & ~ & ~ & ~ & ~ & ~ & ~ & ~ & ~ & ~ & ~ & ~ & ~ & ~ & \checkmark & \checkmark & ~ & \checkmark & ~ & ~ & ~ & ~ \\ \hline
        \multicolumn{28}{l}
        {$\mathcal{A}$: NOMA. $\mathcal{B}$: IRS-assisted. $\mathcal{C}$: Beamforming-optimized. $\mathcal{D}$: Phase-shifts-optimized. $\mathcal{E}$:power-allocator-optimized. $\mathcal{F}$:External Eve. $\mathcal{G}$: Internal Eve. } \\
        \multicolumn{28}{l}
        {$\mathcal{H}$: Neural Networks. $\mathcal{I}$:Secrecy rate.} \\
    \end{tabular} }
\end{table*}

\subsection{ Motivation and Contributions}
%重写版
Due to the inherent openness of electromagnetic signal propagation and reflection, the passive nature of IRS makes it vulnerable to targeted eavesdropping attacks. Coupled with multi-user signal superposition in NOMA, systems are highly susceptible to both external and internal eavesdropping threats. Despite this, there is a conspicuous dearth of research that comprehensively addresses reliable and secure communication in IRS-assisted NOMA networks that are susceptible to both internal and external eavesdropping in the heterogeneous links, posing a complex challenge. Current literature predominantly focuses on optimizing beamforming at BS and phase shifts of IRS to enhance system transmission or security. However, a power allocation policy based on successive interference cancellation (SIC) holds the potential to significantly enhance both transmission efficiency and security, as evidenced in \cite{9594701}. Nonetheless, the joint optimization of the heterogeneous resources, including the IRS phase shift matrix, base station beamforming, and user power allocation policy, remains an area of limited exploration. Additionally, while deep learning algorithms have demonstrated remarkable performance gains in deep feature extraction and parameter optimization, their application in enhancing the security performance of IRS-assisted NOMA networks remains an understudied area, indicating a crucial research gap.

Motivated by the aforementioned context, our objective is to obtain the sum secrecy rate in IRS-assisted NOMA networks. The differences between our work and other existing studies are presented in TABLE I. Our contributions are outlined as follows:

\begin{itemize}
%模型

\item{We propose a heterogeneous secure IRS-assisted NOMA transmission system that addresses both internal eavesdroppers and external eavesdroppers of wireless networks. Our research focuses on planning wireless heterogeneous resource configurations amidst the coexistence of internal and external eavesdroppers.}

\item{We propose the Combinatorial Optimization Graph Neural Networks (CO-GNN) algorithm to design a heterogeneous resource configuration approach. This scheme addresses the beamforming, power allocation, and phase shift problems in our model for security analysis. By directly mapping received signals to optimal parameters, CO-GNN eliminates the necessity for explicit mathematical representations and channel estimation, thereby bolstering resilience against attacks.}

%窃听 安全分析
\item{Simulation results clearly demonstrate that our CO-GNN scheme consistently achieves the highest sum secrecy rate, thereby validating the effectiveness of our the heterogeneous resource configuration approach. Furthermore, as the transmit power increases, as well as the number of IRS reflecting elements or transmit antennas, the security of our model experiences a noteworthy enhancement. Notably, compared with conventional AO schemes, the CO-GNN achieves superior security performance while maintaining low computational complexity. These findings not only establish the superiority of our proposed method but also highlight its practicality and scalability in real-world scenarios.}

\end{itemize}

\subsection{Paper Organization}
The rest of the paper is given as follows. Section II proposes the system model, hardware impairments and problem formulation. Section III gives the designed CO-GNN network for joint optimization. Section IV presents the performance results and the analysis and Section V draws the conclusion.

\section{System Model and Problem Formulation}

\subsection{System Model}
%ABE->IRS->信道系数等等

\begin{figure} % Single column figure
    \includegraphics[width=\linewidth]{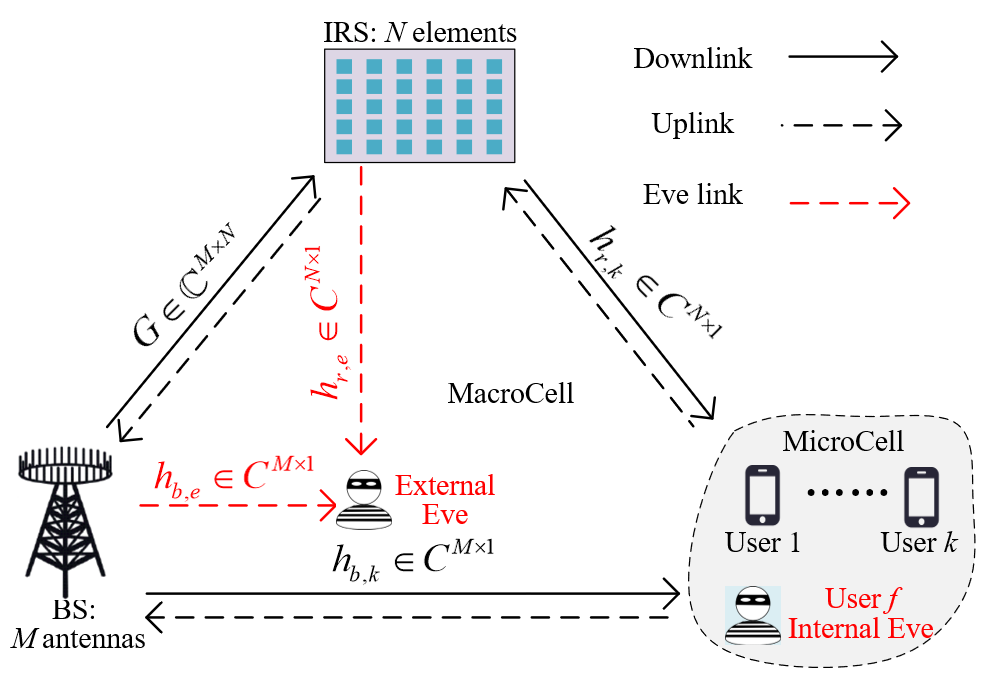}
    \caption{System Model}
	\label{pic:model}
\end{figure}

In this paper, we investigate an IRS-assisted NOMA communication system depicted in Fig. \ref{pic:model}, encompassing downlink, uplink, and eavesdropping links. The uplink and downlink denote signal transmission directions in an IRS-assisted NOMA system: downlink from the base station (BS) or via the IRS to users, and uplink vice versa. Operating under a time division duplex (TDD) system, both the BS and IRS can acquire channel state information (CSI) from users for the downlink. We specifically consider two types of eavesdroppers: external eavesdroppers, unauthorized and not part of NOMA pairing, and internal eavesdroppers within the NOMA pairing, attempting to intercept NOMA signals.

The BS is equipped with $M$ transmit antennas and communicates with $K$ single-antenna users, where $M\ge K$. Eavesdroppers also deploy single antennas. The reflecting IRS is equipped with $N$ low-cost passive reflecting elements capable of digitally controlling phase and amplitude to alter signal propagation directions and effects. Channel gain from the BS to the IRS, $k$-th users, and Eves, and from the IRS to the $k$-th users and Eves are denoted as $\textbf{G} \in \mathbb{C}^{M \times N}$, $\textbf{h}_{b,k} \in \mathbb{C}^{M \times 1}$, $\textbf{h}_{b,e} \in \mathbb{C}^{M \times 1}$, $\textbf{h}_{b,f} \in \mathbb{C}^{M \times 1}$, $\textbf{h}_{r,k} \in \mathbb{C}^{N \times 1}$, $\textbf{h}_{r,e} \in \mathbb{C}^{N \times 1}$, and $\textbf{h}_{r,f} \in \mathbb{C}^{N \times 1}$, respectively. All channel coefficients experience small-scale fading and path loss, modeled by Rayleigh fading. The CSI of users is obtainable at both the BS and IRS but not at eavesdroppers, who remain silent. The downlink channels from the BS to the user $k$ can be expressed as

\begin{equation}
    {\textbf{h}_{b,k}} = {\beta_{0,k}}{\widetilde{\textbf{h}}_{b,k}},
\end{equation}
where ${\widetilde{\textbf{h}}_{b,k}} \sim \mathcal{CN} \left( {0,1} \right)$, and ${\beta _{0,k}}$ denotes the path loss of the downlink.

The IRS is strategically placed to maintain a downlink line-of-sight(LOS) communication pathway with the BS and the users. Therefore, the channels, represented by $\textbf{G}$ and $\textbf{h}_{r,k}$, are characterized using the Rician fading model, expressed as:

\begin{equation}
    \textbf{G} = {\beta _1}\left( {\sqrt {\frac{\kappa }{{1 + \kappa }}} {{\widetilde{\textbf{G}}}^{LOS}} + \sqrt {\frac{\kappa }{{1 + \kappa }}} {{\widetilde{\textbf{G}}}^{NLOS}}} \right),
\end{equation}

\begin{equation}
    {\textbf{h}_{r,k}} = {\beta _{2,k}}\left( {\sqrt {\frac{\kappa }{{1 + \kappa }}} \widetilde{\textbf{h}}_{r,k}^{LOS} + \sqrt {\frac{\kappa }{{1 + \kappa }}} \widetilde{\textbf{h}}_{r,k}^{NLOS}} \right),
\end{equation}
where $\kappa$ signifies the Rician K-factor. The superscripts 'LOS' and 'NLOS' refer to the line-of-sight and non-line-of-sight components, respectively. The terms ${\beta_1}$ and ${\beta_{2,k}}$ denote the path loss from the BS to the IRS, and from the IRS to the $k$-th user, respectively. The elements within the matrices ${{{\widetilde{\textbf{G}}}^{NLOS}}}$ and ${\widetilde{\textbf{h}}_{r,k}^{NLOS}}$ are modeled as independent and identically distributed (i.i.d.) complex Gaussian random variables with zero mean and unit variance, i.e., ${\left[ {{{\widetilde{\textbf{G}}}^{NLOS}}} \right]_{ij}} \sim \mathcal{CN}(0,1)$ and ${\left[ {\widetilde{\textbf{h}}_{r,k}^{NLOS}} \right]_i} \sim \mathcal{CN}(0,1)$.

The LoS component of the channel $\textbf{h}_{r,k}$ depends on the positions of the IRS and users. Let ${\phi _{2,k}}$ and ${\theta_{2,k}}$ represent the azimuth and elevation angles of arrival(AoA) from user $k$ to the IRS. The LoS channel is given by $\widetilde{h}_{r,k}^{LOS} = {\alpha _{IRS}}({\phi _{2,k}},{\theta _{2,k}})$, where the $n$-th element of the IRS steering vector ${\alpha _{IRS}}({\phi _{2,k}},{\theta _{2,k}})$ is defined as \cite{9300189}.

\begin{equation}
    \begin{array}{l}
{\left[ {{\alpha_{IRS}}\left( {{\phi _{2,k}},{\theta _{2,k}}} \right)} \right]_n} = \\
{e^{j\frac{{2\pi {d^{IRS}}}}{{{\lambda _c}}}\left\{ {{i_1}\left( n \right)\sin \left( {{\phi _{2,k}}} \right)\cos \left( {{\theta _{2,k}}} \right) + {i_2}\left( n \right)\sin \left( {{\theta _{2,k}}} \right)} \right\}}}
\end{array},
\label{eq:a_IRS}
\end{equation}
where $d^{IRS}$ represents the distance between two adjacent elements of the IRS, and ${\lambda _c}$ denotes the carrier wavelength. The functions ${i_1}(n) = \bmod(n - 1, 10)$ and ${i_2}(n) = \left\lfloor \frac{n - 1}{10} \right\rfloor$ are defined to facilitate specific calculations. For simplicity, we set $\frac{2{d^{IRS}}}{{\lambda _c}} = 1$.

Let $\left( {{s_k},{y_k},{z_k}} \right)$ denotes the location of the user $k$ and $\left( {{x^{IRS}},{y^{IRS}},{z^{IRS}}} \right)$ represents the location of the IRS, then 

\begin{equation}
    \sin \left( {{\phi _{2,k}}} \right)\cos \left( {{\theta _{2,k}}} \right) = \frac{{{y_k} - {y^{IRS}}}}{{d_k^{IU}}},
    \label{eq:5}
\end{equation}

\begin{equation}
    \sin \left( {{\theta _{2,k}}} \right) = \frac{{{z_k} - {z^{IRS}}}}{{d_k^{IU}}},
    \label{eq:6}
\end{equation}
where ${d_k^{IU}}$ is the distance between IRS and $k$-th user.

Similarly, let ${\phi _0}$ and ${\theta _0}$ denote the azimuth and elevation angles of arrival at the BS. The BS steering vector can then be formulated as
%Similarly, let ${\phi _0}$, ${\theta _0}$ denote the azimuth and elevation angles of arrival to the BS, then the steering vector of the BS can be written as

\begin{equation}
    {\alpha _{BS}}\left( {{\phi _0},{\theta _0}} \right) = \left[ {1, \cdots,{e^{j\frac{{2\pi \left( {M - 1} \right){d^{BS}}}}{{{\lambda _c}}}\cos \left( {{\phi _0}} \right)\cos \left( {{\theta _0}} \right)}}} \right],
\end{equation}
where $d^{BS}$ is the distance between two adjacent BS antennas, and we assume $\frac{{2\pi {d^{BS}}}}{{{\lambda _c}}} = 1$. Let ${\phi _1}$, ${\theta _1}$ denote the azimuth and elevation angles of departure(AoD) from the IRS to the BS, then we can obtain

\begin{equation}
    {\widetilde{\textbf{G}}^{LOS}} = {\alpha _{BS}}\left( {{\phi _0},{\theta _0}} \right){\alpha _{IRS}}{\left( {{\phi _1},{\theta _1}} \right)^H}.
    \label{eq:GLos}
\end{equation}

%Given the location of BS $\left( {{x^{BS}},{y^{BS}},{z^{BS}}} \right)$ and the location of the IRS $\left( {{x^{IRS}},{y^{IRS}},{z^{IRS}}} \right)$, we have

From (\ref{eq:5}) and (\ref{eq:6}), we can get

\begin{equation}
    \cos \left( {{\phi _0}} \right)\cos \left( {{\theta _0}} \right) = \frac{{{x^{IRS}} - {x^{BS}}}}{{{d^{BI}}}},
\end{equation}

\begin{equation}
    \sin \left( {{\phi _1}} \right)\cos \left( {{\theta _1}} \right) = \frac{{{y^{BS}} - {y^{IRS}}}}{{{d^{BI}}}},
    \label{eq:sincos_phi1}
\end{equation}

\begin{equation}
    \sin \left( {{\theta _1}} \right) = \frac{{{z^{BS}} - {z^{IRS}}}}{{{d^{BI}}}},
    \label{eq:sin_1}
\end{equation}
where $d^{BI}$ is the distance between the IRS and the BS.

Let $s_k \in \mathbb{C}$ be the signal intended for transmission from BS to $k$-th user. The BS employs a beamforming strategy,  utilizing a beamforming vector $\textbf{w}_k\in \mathbb{C}^M$, and encodes the signals for all users simultaneously with the power allocation factor $\textbf{a}_k$. Both the beamforming and power allocation factor must satisfy some constraints: $\sum\nolimits_{k = 1}^K {{{\left\| {{\textbf{w}_k}} \right\|}^2}}  \le {P_t}$ and $\sum\nolimits_{k = 1}^K {{\textbf{a}_k}}  = 1$, where $P_t$ is the transmission power at BS. The reflection matrix of IRS is denoted as $\mathbf{\Phi}=diag\left\{ {{e^{j{\theta _1}}}, \cdots ,{e^{j\theta n}}, \cdots ,{e^{j{\theta _N}}}} \right\}$, where $\theta_n$ is the phase shifts at $n$-th reconfiguration element. So the received signal at $k$-th user is denoted as

%The signal is sent directly or through the IRS indirectly to the users. 

\begin{equation}
    {\textbf{y}_k} = \sum\limits_{i = 1}^K {{{\left( {{\textbf{h}_{b,k}} + {\textbf{G}}\mathbf{\Phi} {\textbf{h}_{r,k}}} \right)}^{\rm T}}} {\textbf{w}_k}\sqrt {{P_t}{\textbf{a}_k}} {s_k} + {n_k},
    \label{eq:1}
\end{equation}
%where the $P_{BS}$ is the transmission power for all users at BS in NOMA networks. To guarantee user fairness, $\textbf{w}_k$ represents the power allocation factor of users $k$, which satisfies that $w_p+w_q=1$. And it's noting that user $Q$ has better channel qualification than $P$ since we stipulate that $P$ is closer to the IRS and BS. %对于信道状态，根据代码来
where the $n_k$ denotes the additive white gaussian noise (AWGN) at user $k$ with mean power parameter ${\sigma ^2}$.

In NOMA systems, users employ the SIC technique for signal decoding. The demodulation order in SIC is established based on the combined channel's equivalent channel gain. We adhere to the principle that the user with the strongest channel gain is demodulated first, while the users with weaker channel gains receive information directly. The users are ordered as ${\left\| {{\textbf{h}_1}} \right\|^2} \ge {\left\| {{\textbf{h}_2}} \right\|^2} \ge  \cdots  \ge {\left\| {{\textbf{h}_K}} \right\|^2}$, where the $k$-th ordered users corresponds to the $k$-th Strongest channels. For the $k$-th strongest user, it decodes its signal by treating other weaker users' signals as interference. The signal-to-interference-plus-noise ratio (SINR) for the strongest $k$-th user decoding its signal can be expressed as

 %The users in NOMA systems employ the SIC technique to decode the signals. Specifically, the SIC demodulation order is determined by the equivalent channel gain of the joint channel, and the higher the equivalent channel gain, the lower the SIC demodulation order. For $k$-th user, after cancellingthe interference signals from all weaker users, it decodes its own signal by treating the signals from the other users as interference. The Signal to Interference plus Noise Ratio (SINR) for $k$-th user decoding its own signal is expressed as

%According to NOMA protocol, the SIC scheme is introduced at the stronger user $P$ to first decode the weaker user $Q$'s information and then decode its own signal. In this case, the received SINR at user $P$ can be shown as

\begin{equation}
    {\gamma _k} = \frac{{{\textbf{a}_k}{P_t}{{\left| {{{\left( {{\textbf{h}_{b,k}} + {G}\mathbf{\Phi} {\textbf{h}_{r,k}}} \right)}^{\rm T}}{\textbf{w}_k}} \right|}^2}}}{{\sum\limits_{i = k + 1}^K {{\textbf{a}_i}{P_t}{{\left| {{{\left( {{\textbf{h}_{b,k}} + {\textbf{G}}\mathbf{\Phi} {\textbf{h}_{r,k}}} \right)}^{\rm T}}{\textbf{w}_i}} \right|}^2}}  + \sigma _k^2}}.
\end{equation}

%Then user $Q$ will treat user $P$'s signal as interference and directly acquire its signal. Hence, the received SINR at user $Q$ can be obtained as

%\begin{equation}
    %{\gamma _q} = \frac{{{w_q}P{{\left| {{\textbf{h}_{bq}} + {\textbf{h}_{br}}\mathbf{\Phi} \textbf{h}_{rq}^H} \right|}^2}}}{{{\sigma ^2}}}
%\end{equation}

Consequently, the transmission rate for $k$-th user can be denoted as

\begin{equation}
    {R_k} = {\log _2}\left( {1 + {\gamma _k}} \right).
    \label{eq:r_k}
\end{equation}

%Likewise, the transmission rate for user $Q$ can be written as

%\begin{equation}
    %{R_q} = \log \left( {1 + \frac{{{w_q}P{{\left| {{\textbf{h}_{bq}} + {\textbf{h}_{br}}\mathbf{\Phi} \textbf{h}_{rq}^H} \right|}^2}}}{{{\sigma ^2}}}} \right)
%\end{equation}

Considering the detrimental effects of the intricate electromagnetic environment on secure communication via exposed radio signals, we examine both external and internal wiretapping scenarios in the heterogeneous links to assess the secrecy performance of IRS-NOMA networks.

%Due to the adverse impact of the complex electromagnetic environment on secure communication through exposed radio signals, we consider both external and internal wiretapping scenarios to evaluate the secrecy performance of RIS-NOMA networks. 

An external eavesdropper is unauthorized and not part of the NOMA pairing. The broadcast nature of wireless networks increases the likelihood of unauthorized eavesdroppers intercepting NOMA signals.

1) External Eve: the external Eve is unauthorized and not part of the NOMA pairing. The broadcast nature of wireless networks increases the likelihood of unauthorized eavesdroppers intercepting NOMA signals. To this end, the received signal at Eve can be derived as

\begin{equation}
    {\textbf{y}_{EE}} = \sum\limits_{i = 1}^K {{{\left( {{\textbf{h}_{b,e}} + {\textbf{G}}\mathbf{\Phi} {\textbf{h}_{r,e}}} \right)}^{\rm T}}} {\textbf{w}_k}\sqrt {{P_t}{\textbf{a}_k}} {s_k} + {n_e},
\end{equation}
where the $n_e$ denotes the AWGN at Eve. Referring to the analysis above, the external Eve can decode the information of user $k$ by applying SIC. So the SINR for $E$ to wiretap the signal from user $k$ can be established as

\begin{equation}
    {\gamma _{e \to k}} = \frac{{{\textbf{a}_k}{P_t}{{\left| {{{\left( {{\textbf{h}_{b,e}} + {\textbf{G}}\mathbf{\Phi} {\textbf{h}_{r,e}}} \right)}^{\rm T}}{\textbf{w}_k}} \right|}^2}}}{{\sum\limits_{i = k + 1}^K {{\textbf{a}_i}{P_t}{{\left| {{{\left( {{\textbf{h}_{b,e}} + {\textbf{G}}\mathbf{\Phi} {\textbf{h}_{r,e}}} \right)}^{\rm T}}{\textbf{w}_i}} \right|}^2}}  + \sigma _k^2}}.
    \label{eq:gamma_ek}
\end{equation}

Then the transmission rate for external eve can be obtained like (\ref{eq:r_k}).
%\begin{equation}
%    {R_{e \to k}} = {\log _2}\left( {1 + {\gamma _{e \to k}}} \right).
%\label{secrecy_ex}
%\end{equation}

2) Internal Eve: In this scenario, although the distant user $f$ is a legitimate user, its weaker channel condition urges it to inadvertently overhear transmissions intended for other legitimate users, thus functioning as an internal eavesdropper. Consequently, the signal received at internal Eve $f$ can be expressed as

\begin{equation}
    {\textbf{y}_{IE}} = \sum\limits_{i = 1}^{K-1} {{{\left( {{\textbf{h}_{b,f}} + {\textbf{G}}\mathbf{\Phi} {\textbf{h}_{r,f}}} \right)}^{\rm T}}} {\textbf{w}_k}\sqrt {{P_t}{\textbf{a}_k}} {s_k} + {n_f}.
\end{equation}

At this moment, the SINR $\gamma _{f \to k,k \neq f}$ for internal eavesdropper $f$ can be derived like (\ref{eq:gamma_ek}) and the transmission rate for $f$ also can be obtained like (\ref{eq:r_k}).

%\begin{equation}
%    {\gamma _{f \to k,k \neq f}} = \frac{{{\textbf{a}_k}{P_t}{{\left| {{{\left( {{\textbf{h}_{b,f}} + {\textbf{G}}\mathbf{\Phi} {\textbf{h}_{r,f}}} \right)}^{\rm T}}{w_f}} \right|}^2}}}{{\sum\limits_{i = f + 1}^K {{\textbf{a}_i}{P_t}{{\left| {{{\left( {{\textbf{h}_{b,f}} + {\textbf{G}}\mathbf{\Phi} {\textbf{h}_{r,f}}} \right)}^{\rm T}}{\textbf{w}_i}} \right|}^2}}  + \sigma _f^2}}.
%\end{equation}

%Then the transmission rate for internal eve can be denoted as
%\begin{equation}
%    {R_{f \to k,k \neq f}} = {\log _2}\left( {1 + {\gamma _{f \to k}}} \right).
%\end{equation}

According to the definition of the secrecy rate,  for each user, the secrecy rate $R_k^{sec}$ can be expressed as

\begin{equation}
    \begin{array}{*{20}{c}}
{R_k^{\sec } = {{\left[ {{R_k} - {R_{\varphi \to k}}} \right]}^ + }}&{\varphi = e,f}.
\end{array}
\label{secrecy_in}
\end{equation}
%where the eavesdropping rate can refer to either an internal eavesdropping rate or an external eavesdropping rate.

%Our objective is to jointly optimize the robust BS's power distribution matrix $\textbf{w}$ based on NOMA protocol and the robust IRS's phase shift matrix $v$ to maximize the secrecy rate $R_k^{sec}$. By mapping the received pilots directly to the optimized power distribution and phase shifts for utility maximization, we can bypass the channel estimation getting the optimized transmission strategy more efficiently.

%As a consequence, we can get the optimal power distribution $\textbf{w}$ and phase shifts $v$ based directly on the received pilots $Y$. To this end, our goal can be formulated as

%$R_k^{sec}$

\subsection{Problem Formulation}
Our objective is to jointly optimize the robust BS's beamforming $\textbf{w}$ and power allocation matrix $\textbf{a}$ based on NOMA protocol and the robust IRS's phase shift matrix $\mathbf{\Phi}$ to maximize the sum secrecy rate. By mapping the received signals directly to the optimized beamforming, power allocation and phase shifts for utility maximization, we can bypass the channel estimation getting the optimized transmission strategy more efficiently.

%由于硬件损伤不影响传输速率与安全速率，则硬件损伤下的目标函数可与无硬件损伤情况一同考虑。Since hardware impairments do not affect the transmission rate and the secure rate, the objective function under hardware impairments can be considered alongside the scenario without hardware impairments.
As a consequence, we can get the optimal beamforming $\textbf{w}$, power allocation $\textbf{a}$ and phase shifts $\mathbf{\Phi}$ based directly on the received signals $Y$ by a function $g(\cdot)$. 
% Since hardware impairments do not affect the transmission rate and the secrecy rate, the objective function under hardware impairments can be considered analogous to the scenario without hardware impairments. To this end, our optimization goal can be formulated as

\begin{equation}
\begin{array}{l}
\begin{array}{*{20}{c}}
{\mathop {maximize}\limits_{(\textbf{w},\textbf{a},\mathbf{\Phi} ) = g(\textbf{Y})} }&{\sum\limits_{k = 1}^K {R_k^{\sec }\left( {\textbf{w},\textbf{a},\mathbf{\Phi} } \right)} }
\end{array}\\
\begin{array}{*{20}{c}}
{s.t.}&{\begin{array}{*{20}{c}}
{\begin{array}{*{20}{c}}
{\sum\nolimits_{k = 1}^K {{{\left\| {{\textbf{w}_k}} \right\|}^2}}  \le {P_t}}  \vspace{1ex}\\
{\sum\nolimits_{k = 1}^K {{\textbf{a}_k}}  = 1}
\end{array}}\vspace{1ex}\\
{\begin{array}{*{20}{c}}
{\left| {{e^{j{\theta _n}}}} \right| = 1,0 \le {\theta _n} \le 2\pi }&{\forall n}
\end{array}}\vspace{1ex}\\
{\begin{array}{*{20}{c}}
{{{\left\| {{\textbf{h}_{b1}} + {\textbf{h}_{br}}\mathbf{\Phi} {\textbf{h}_{r1}}} \right\|}^2} \ge {{\left\| {{\textbf{h}_{b2}} + {\textbf{h}_{br}}\mathbf{\Phi} {\textbf{h}_{r2}}} \right\|}^2}}\vspace{1ex}\\
{ \ge  \cdots  \ge {{\left\| {{\textbf{h}_{bK}} + {\textbf{h}_{br}}\mathbf{\Phi} {\textbf{h}_{rK}}} \right\|}^2}}
\end{array}}\vspace{1ex}\\
{\begin{array}{*{20}{c}}
{R_{sec} \ge R_{min}}
\end{array}}
\end{array}}
\end{array}
\end{array}
\label{eq:goal}
\end{equation}

%where $g(\cdot)$ represents a function that maps the received signals to the beamforming vector $\textbf{w}$, power distribution vector $\textbf{a}$, and phase shifts matrix $\mathbf{\Phi}$. 

The constraint $R_{sec} \ge R_{min}$ guarantees that all legitimate users achieve a secure transmission rate no less than the predefined threshold $R_{min}$. This design effectively prevents users with poor channel conditions from being sacrificed due to unbalanced resource allocation, thereby significantly enhancing the system's service fairness. Specifically, when the equivalent channel gain of a user is detected to be low, the CO-GNN algorithm dynamically adjusts its beamforming vector $\textbf{w}_k$ and power allocation factor $\textbf{a}_k$ to prioritize compensating its secrecy rate until the $R_{min}$ requirement is satisfied.

Due to the non-convex nature of the objective function in the problem (\ref{eq:goal}), solving it computationally presents a significant challenge. To address this issue effectively and circumvent the need for channel estimation, we have developed a deep neural network to visualize the mapping function $g(\cdot)$ and to train the heterogeneous network parameters using the received signals to achieve best sum secrecy rate.

\section{Proposed Deep Learning Framework}

\begin{figure*} % Single column figure
	\includegraphics[width=\linewidth]{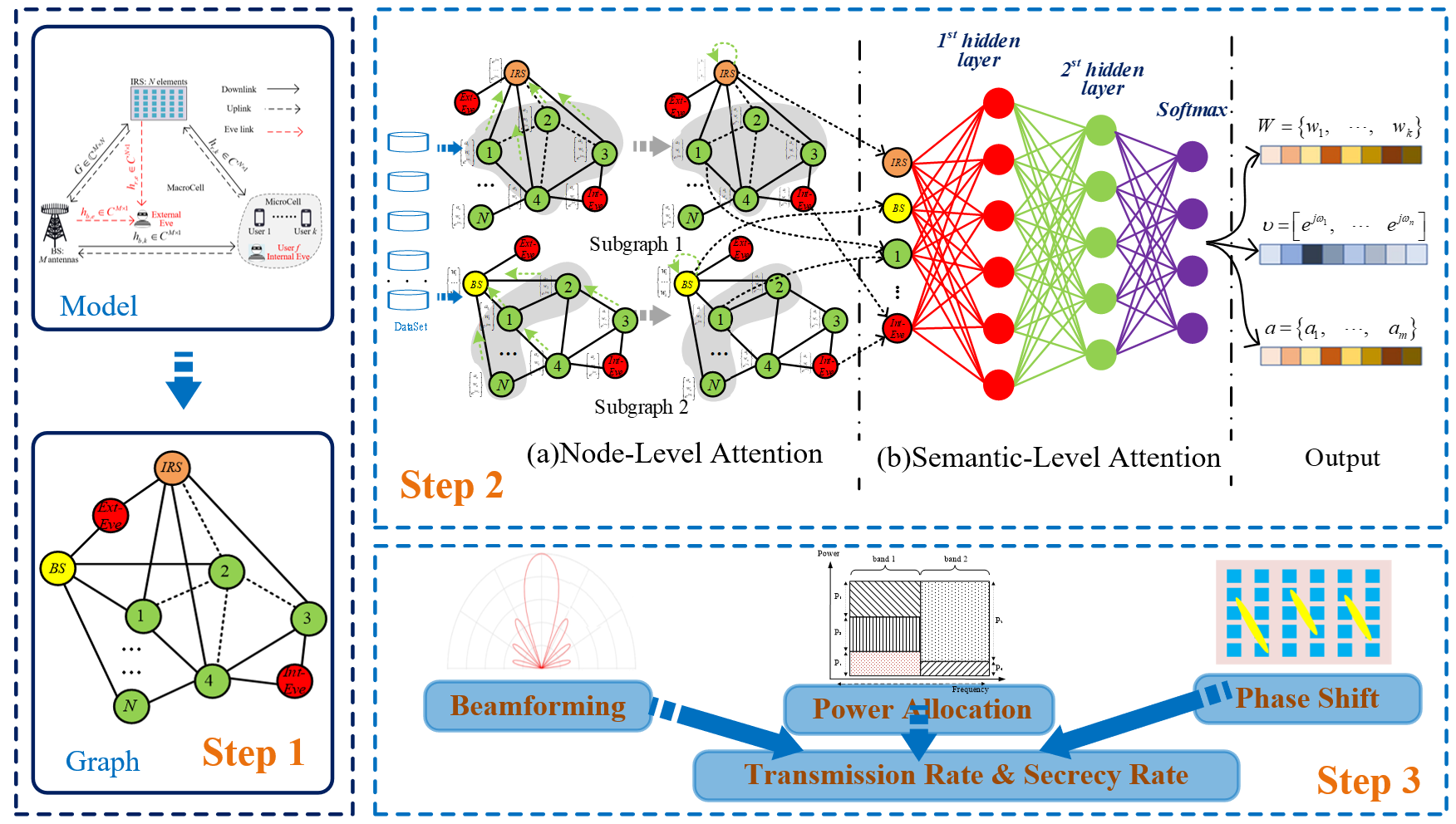}
    \caption{Proposed Deep Learning Framework: CO-GNN}
	\label{pic:GNN}
\end{figure*}

%使用目标函数的倒数为网络的损失函数，我们可以通过目标函数对恶意窃听者的抑制，联合优化波束赋形，相移矩阵以及功率分配，使得信号向着利于合法用户的方向发射，并抑制窃听者的窃听。

Solving the optimization problem delineated in (\ref{eq:goal}) poses significant challenges due to its non-convex nature. Traditional optimization methods often encounter computational hurdles, particularly in scenarios requiring channel estimation. Consequently, applying these conventional techniques to devise an efficient secure joint optimization policy in dynamically changing environments is generally impractical. 
In response to this optimization challenge posed by (\ref{eq:goal}), we introduce a neural network, CO-GNN, to model the mapping function $g(\cdot)$. By employing the proposed CO-GNN, we achieve joint optimization and get the beamforming, power allocation, and phase shifts straightly from the received signals, ultimately maximizing the secrecy network utility. 
By leveraging the opposite number of the objective function as the loss function for our network, we are able to achieve the joint optimization through the suppression of malicious eavesdroppers, as dictated by the objective function, thereby ensuring that the signal is transmitted in a direction that favors legitimate users while simultaneously mitigating eavesdropping attempts. The comprehensive framework of this approach is depicted in Fig. \ref{pic:GNN}. In the following, we elaborate on the graphical representation, our CO-GNN, and the training procedure.

\subsection{Graphical Presentation of Users and IRS}
%情况总述，主要是有用户见干扰所以用GNN，还需要更有力的理由，如何用图表示
%为什么要用图和GNN，要与传统优化方法和CNN两方面对比，所以还需要改下调整下

To precisely characterize the interplay between nodes and IRS and minimize mutual interference among users, we employ graphs to represent users and IRS, thereby capturing the underlying relationships and mitigating negative interference. To achieve this, we devise a combinatorial optimization graph neural network to jointly optimize beamforming $\textbf{w}$, power allocation factors $\textbf{a}$ at BS and the phase shifts $\mathbf{\Phi}$ at the IRS, aiming for superior system performance. Our CO-GNN is apt for communication networks due to its robust learning capacity in capturing interactions between users and IRS and extracting spatial information embedded in the network topology. Moreover, our CO-GNN maintains permutation invariance and permutation equivariance properties for graphs. Specifically, regardless of permutations or shuffles in the input index labels, the GNN network can still generate the correspondingly changing or unchanging output. Here, permutation invariance ensures that phase shifts are agnostic to the ordering of user channels, while permutation equivariance implies that if user channels are rearranged, the beamforming and power allocation vectors will be similarly permuted. The parameters of the GNN can be shared among users, facilitating its generalization to scenarios with varying user counts. Furthermore, the GNN architecture reduces model complexity compared to fully connected neural networks.  

We employ a graph to model the interactions between the IRS and user nodes. The IRS is represented by node $0$, while the $K$ users are represented by nodes $1$ through $K$. Meanwhile, the information of Node $k$ is associated with a representation vector, denoted as $r_k$, where $k=0,1,...,K$. Within our CO-GNN, the vectors are fed and updated layer by layer, with each update considering the vectors from the previous layer as input. After multiple layers of processing, the vector of each node will contain sufficient information to facilitate the design of beamforming $\textbf{w}$, power allocator $\textbf{a}$, and phase shits $\mathbf{\Phi}$. Specifically, the update of each user node's vector is a function of all its neighboring user nodes' vectors as well as the IRS node's vector. This design enables our CO-GNN to learn how to mitigate interference by capturing the relationships between neighboring nodes and the IRS as well as getting the joint optimization results for improved secrecy performance. The update of the IRS node is a function of all the user nodes, which enables our CO-GNN to learn to configure the phase shifts on all the channels.

\subsection{Combinatorial Optimization Graph Neural Networks}

We employ CO-GNN to train and jointly optimize the heterogeneous resource configurations, including the beamforming vector $\textbf{w}$, the power allocation vector $\textbf{a}$, and the phase shifts $\mathbf{\Phi}$. Our CO-GNN specifically focuses on learning the graph representation vector $r_k$ through a multi-layer architecture consisting of an input layer, two message passing layers, and an output layer. The network ingests the received signals from the user node of the graph, which serve as initial input values for joint optimization. These inputs are then propagated through two message passing layers, which contain the effective aggregation and combination functions, resulting in jointly optimized representations. Finally, an output layer straightly transforms and maps the representations into the optimized beamforming vector $\textbf{w}$, power allocation vector $\textbf{a}$, and phase shifts $\mathbf{\Phi}$ via a fully connected layer. The overall architecture of our CO-GNN is depicted in Fig. \ref{pic:GNN}.
%denoted as ${\rm{r}}_k^0$

1)Input Layer: The input layer receives the feature vectors from the received signals for each user in the graph, denoted as $\textbf{Y}_k$ for $k=1,..., K$. This is attributed to the fact that the received signal encapsulates ample information pertaining to beamforming, power allocation, and phase shifts, rendering it a valuable input for achieving enhanced outcomes through joint optimization of our CO-GNN. Given that the signal comprises both real and imaginary components, for the user nodes, we can derive the updated result ${\rm{r}}_k^1$ as follows:

\begin{equation}
    {\rm{r}}_k^1 = f_{MLP}^1\left( {{{\left[ {\left( {{\textbf{Y}_k}} \right).real} \right]}^T},{{\left[ {\left( {{\textbf{Y}_k}} \right).imag} \right]}^T}} \right),
\end{equation}
where $f_{MLP}^1\left(  \cdot  \right)$ denotes a multilayer perceptron (MLP) that comprises two fully connected hidden layers.

For the IRS node, since all signals are reflected via the IRS, it is crucial to ensure that the received signals all contain an equivalent amount of information pertaining to the IRS. Consequently, the $\rm{r}_0^1$ can be derived as the mean of all the signals received from the user nodes:

\begin{equation}
 r_0^1 = f_{MLP}^1\left( {\frac{1}{K}\sum\limits_{k = 1}^K {\left[ {{{\left[ {\left( {{\textbf{Y}_k}} \right).real} \right]}^T},{{\left[ {\left( {{\textbf{Y}_k}} \right).imag} \right]}^T}} \right]} } \right).
\end{equation}

2) Message Passing Layers:  The message passing layers are the pivotal layers in our CO-GNN for joint optimization. These layers leverage neighbor aggregation and combination strategies to extract information from nodes, enabling the generation of a representation vector encompassing beamforming, power allocator and phase shifts for joint optimization. The representation vector are derived by aggregating the characteristics of neighboring nodes and subsequently combining them, effectively amounting to a message passing process \cite{2018How}. In our implementation of the message passing layer, the aggregation function serves to compile the features of neighboring nodes into a message vector, which can then be relayed to the central node. Conversely, the combination function updates the node's representation at the current instant, integrating both the current representation of the node and the messages garnered from the aggregation function. A crucial aspect in the design of GNNs lies in the selection of a suitable aggregation function $f_{aggregate}\left(  \cdot  \right)$ and a combination function $f_{combine}\left(  \cdot  \right)$ that enable the GNN to scale effectively and jointly optimize well. As for user nodes, we choose the aggregation function as

\begin{equation}
    f_{aggregate}^i\left( {{{\left\{ {r_j^{i - 1}} \right\}}_{j \in N\left( k \right)}}} \right) = \Theta \left( {{{\left\{ {f_{nn}^i\left( {r_j^{i - 1}} \right)} \right\}}_{j \in N\left( k \right)}}} \right),
\end{equation}
where $i$ is $i$-th aggregate layer, $i=1,..,I$, $N(k)$ denotes the set of neighboring nodes of the node $k$, $\Theta$ is the mean-pooling function, and $f_{nn}^i$ is a fully connected hidden layer. This aggregation function serves the purpose of gathering information from neighboring nodes and consolidating it for transmission to the central node. Mean-pooling ensures that interference features from all users are equally considered, thereby preventing resource allocation skewness. Its low-variance property not only enhances the model's robustness in dynamic channel conditions but also provides a smooth optimization trajectory for secrecy rate maximization.

As for the user combination function, it integrates the current node representation and the aggregated neighbor messages, as discussed in \cite{2018How}, and is realized through the use of a MLP:

\begin{equation}
	f_{combine}^i\left( {\left\{ {r_k^{i - 1}} \right\}} \right) = f_{MLP}^i\left( {r_k^{i - 1},f_{aggregate}^i\left( {{{\left\{ {r_j^{i - 1}} \right\}}_{j \in N\left( k \right)}}} \right)} \right).
\end{equation}
	
With the role of aggregation and combination, the message passing layer for the user can be obtained as

\begin{equation}
	    r_k^i = f_{MLP}^i\left( {f_{MLP}^i\left( {r_0^{i - 1}} \right),f_{combine}^i\left( {\left\{ {r_k^{i - 1}} \right\}} \right)} \right),	
\end{equation}
where the IRS node representation vector $r_0^{i}$ can be present as

\begin{equation}
    r_0^i = f_{MLP}^i\left( \begin{array}{l}
	f_{MLP}^i\left( {r_0^{i - 1}} \right),\\
	\Delta \left( {f_{MLP}^i\left( {r_1^{i - 1}} \right), \cdots ,f_{MLP}^i\left( {r_K^{i - 1}} \right)} \right)
	\end{array} \right),
\end{equation}
where $\Delta$ represents the max-pooling function, known for its empirical performance and its correspondence to the prevailing notion that multiuser interference is predominantly influenced by the strongest user. Max-pooling pinpoints the user most susceptible to the eavesdropping link and amplifies its feature representation. This compels the IRS phase shifts to prioritize suppressing signal leakage from this user, thereby optimizing the system's secrecy rate. %这里要不要引用论文

3) Output layer: After the completion of $I$ message passing layers, we can generate the joint optimized phase shifts $\mathbf{\Phi} \in {\mathbb{C}^N}$, the beamforming $\textbf{w} \in {\mathbb{C}^{M \times K}}$, and the power allocation factors $\textbf{a} \in {\mathbb{R}^K}$ via the final output layer. Specifically, through the fully connected layer of corresponding size, the power allocation and beamforming are derived from the representation vectors of the user nodes, while the phase shifts are derived from the representation vectors of the IRS nodes.

When dealing with the phase shifts, as we initially separated the real and imaginary parts at the input layer, it is necessary at this stage to combine them to obtain the phase shifts in complex form.
	
\begin{equation}
{r_\mathbf{\Phi}} = \left[ {f_{nn}^{out}\left( {r_0^I} \right)\left( {1:N} \right),f_{nn}^{out}\left( {r_0^I} \right)\left( {N + 1:2N} \right)} \right] \in {\mathbb{R}^{N \times 2}},
\end{equation}

\begin{equation}
	    {\mathbf{\Phi}_n} = \frac{{{{\left[ {{r_\mathbf{\Phi}}} \right]}_{n,1}}}}{{\sqrt {\left[ {{r_\mathbf{\Phi}}} \right]_{n,1}^2 + \left[ {{r_\mathbf{\Phi}}} \right]_{n,2}^2} }} + j\frac{{{{\left[ {{r_\mathbf{\Phi}}} \right]}_{n,2}}}}{{\sqrt {\left[ {{r_\mathbf{\Phi}}} \right]_{n,1}^2 + \left[ {{r_\mathbf{\Phi}}} \right]_{n,2}^2} }}.
\end{equation}

For the beamforming and power allocator, the representation vectors of user nodes are needed to through the fully connected layer.	
\begin{equation}
	{r_w} = \left[ {f_{nn}^{out}\left( {r_1^I} \right), \cdots ,f_{nn}^{out}\left( {r_K^I} \right)} \right] \in {R^{2M \times K}},
	\label{eq:rw}
\end{equation}

\begin{equation}
    \textbf{w} = \frac{{{r_w}}}{{\left\| {{r_w}} \right\|}}\left( {1:M,:} \right) + j\frac{{{r_w}}}{{\left\| {{r_w}} \right\|}}\left( {M + 1:2M,:} \right).
\end{equation}

Similarly, we can also get the power allocation vector $\textbf{a}$ like (\ref{eq:rw}) which satisfies $\textbf{a}_k^{out} \in [0,1]$ and $\sum\limits_1^K {\textbf{a}_k^{out}}  = 1$.

\subsection{The Training}

%在每轮训练中，损失不断向网络反馈，让训练的波束赋形等向着提高合法用户的传输速率，降低窃听者速率的方向优化，损失不断减小，而安全和速率不断增大，最终得到联合优化好的波束赋形等及最好的安全和速率。
The loss function of CO-GNN is defined as the negative value of the sum secrecy rate:
\begin{equation}
    \mathcal{L} = - \sum_{k = 1}^{K} R_{k}^{\mathrm{sec}}(\mathbf{w}, \mathbf{a}, \mathbf{\Phi})
\end{equation}

The optimization process can be formulated as a minimization problem for this non-convex function. Despite the overall non-convexity of the loss function, convergence guarantees are achieved through the following design principles:

First, local convergence of gradient descent. Assuming the learning rate $\eta$ satisfies the Lipschitz continuity condition, i.e., there exists a constant $\textit{L}>0$ such that:

\begin{equation}
    \left\lVert \nabla \mathcal{L}(\Theta^{(t + 1)}) - \nabla \mathcal{L}(\Theta^{(t)}) \right\rVert \leq L \left\lVert \Theta^{(t + 1)} - \Theta^{(t)} \right\rVert
\end{equation}
where $\Theta$ denotes the model parameters， and the adaptive learning rate adjustment via the Adam optimizer ensures monotonic decrease of the loss function during iterations and convergence to a local minimum.

Second, smoothness of the loss function. Since the message-passing layers in CO-GNN employ continuously differentiable MLPs and pooling operations, the gradient of loss $\mathcal{L}$ with respect to model parameters $\Theta$ always exists and is computable, satisfying:
\begin{equation}
    \nabla{\Theta} \mathcal{L} = - \sum_{k = 1}^{K} \nabla{\Theta} R_{k}^{\mathrm{sec}},
\end{equation}

This guarantees the validity of gradient descent updates.

% Notably, the loss function is formulated as the negation of the sum secrecy rate, which is defined as the rate difference between legitimate users and eavesdroppers, as specified in (\ref{eq:goal}).
The proposed training algorithm for CO-GNN is outlined in Algorithm \ref{alg:alg1}.  In each training iteration, the loss is iteratively fed back to the network, guiding the optimization of beamforming, power allocation, and phase shifts in a manner that maximizes the transmission rate for legitimate users while minimizing the eavesdropping rate. As the loss value gradually diminishes, the system's sum secrecy rate performance increase concurrently, culminating in the achievement of jointly optimized parameters and the optimal sum secrecy rate. Moreover, the scenarios involving external and internal eavesdroppers are designated as $E_{EX}$ and $E_{IN}$, respectively.

\begin{algorithm}[H]
%\algsetup{linenosize=\small} \scriptsize
\renewcommand{\algorithmicrequire}{ \textbf{Input:}}
\renewcommand{\algorithmicensure}{ \textbf{Output:}}
\caption{CO-GNN Joint Optimization Training Algorithmn}\label{alg:alg1}
\begin{algorithmic}[1]
\REQUIRE Graph $\mathbb{G}$, input signals $\textbf{Y}$, channel group $\textbf{H}$, designed aggregation functions $AGGREGATE_i$, designed combination functions $COMBINE_i$, input layer $MLP_{IN}$, output layer $MLP_{OUT}$
\ENSURE Optimized $\textbf{w}$, $\textbf{a}$, $\mathbf{\Phi}$
%Trained graph neural network $G$

Initialize 
\STATE $\textbf{H}_{R}$, $\textbf{H}_{I}$ $\leftarrow$ Channel2real(\textbf{H})
\REPEAT
\FOR{each episode}
\STATE $\textbf{Y}_{ini} = $ 0
\FOR{$k \in K$}
\STATE $r_{U} \leftarrow MLP_{IN}(\textbf{Y}[k])$
\STATE $\textbf{Y}_{ini} =  \textbf{Y}_{ini}+Y[k]$ 
\ENDFOR
\STATE $Input_{\mathbf{\Phi}} \leftarrow  Mean(\textbf{Y}_{ini}) $
\STATE $r_{\mathbf{\Phi}} \leftarrow  MLP_{IN}(Input_{\mathbf{\Phi}}) $
\FOR{$i \in I$}
\STATE $r_{U}, r_{\mathbf{\Phi}} \leftarrow AGGREGATE_i(r_{U}, r_{\mathbf{\Phi}})$
\STATE $r_{U}, r_{\mathbf{\Phi}} \leftarrow COMBINE_i(r_{U}, r_{\mathbf{\Phi}})$
\ENDFOR
\STATE $\textbf{w}, \textbf{a}, \mathbf{\Phi} \leftarrow MLP_{OUT}(r_{U}, r_{\mathbf{\Phi}})$
\STATE $E_{IN}$: Calculate loss function $L_{IN}(\textbf{H}_R,\textbf{H}_I,\textbf{w},\textbf{a},\mathbf{\Phi})$ using (\ref{secrecy_in})
\STATE $E_{EX}$: Calculate loss function $L_{EX}(\textbf{H}_R,\textbf{H}_I,\textbf{w},\textbf{a},\mathbf{\Phi})$
\STATE Update network weights
\ENDFOR
\UNTIL reaches the maximum training process or the loss does not decreased over $30$ consecutive epochs %30是否要替换为符号
\RETURN Optimized $\textbf{w}$, $\textbf{a}$, $\mathbf{\Phi}$

\end{algorithmic}
\end{algorithm}

\section{Performance Results and Analysis}

In this section, we present a performance evaluation for the proposed CO-GNN to solve the sum secrecy rate optimization problems in IRS-assisted NOMA networks.
%provide the numerical results of the GNN algorithms for solving the secrecy rate optimization problems in RIS-assisted NOMA networks.

\subsection{Setting and Benchmarks}

%table
\begin{table}[!ht]
\renewcommand\arraystretch{1.2}
    \centering
    \caption{Simulation Parameters}\label{table:para}
    \begin{tabular}{l l}
    \hline
    \hline
        \rule{0pt}{10pt}System Parameter & Numerical Value \\ \hline
        Number of antennas & 5 \\ 
        Number of reflecting elements & 100 \\
        Number of users & 2 \\  
        Number of eavesdropper & 1 \\
        Number of message passing la & 1 \\
        The Rician factor & 10dBm \\ 
        Transmit Power & 30dBm \\ 
        Noise power  & -100dBm \\ 
        The coordinates of BS & (0,0,0) \\ 
        The coordinates of IRS & (rand(20,30),rand(20,30),0) \\ 
        The coordinates of users & (rand(30,50),rand(30,50),-10) \\ 
        The coordinates of external eve & (rand(50,100),rand(30,50),-20) \\ 
        Path loss for $\textbf{h}_{d,k}$ & 32.6+36.7lgd \\ 
        Path loss for \textbf{G} and $\textbf{h}_{r,k}$ & 30.0+22.0lgd \\ 
        Learning rate & $1  \times 10^{-4}$ \\ 
        Max epochs & 100 \\ 
        No-increasing epochs & 30 \\ 
        Training samples & 10000 \\ \hline
    \end{tabular}
\end{table}

The program in this study was developed based on the TensorFlow\cite{Mart2015TensorFlow} framework, and all experiments were conducted using an RTX 3090 Ti GPU. And the hyperparameters used in our model are shown in TABLE \ref{table:para}. Specifically, our neural network was trained to utilize the Adam optimizer \cite{2014Adam} with an initial learning rate $1 \times 10^{-4}$. The training process is designed to terminate if it reaches the maximum of $100$ epochs or if the loss function does not decrease over $30$ consecutive training epochs. During each training epoch, we iterate $100$ times to update the parameters of the neural network, with $10000$ training samples being used to compute the gradients in each iteration. In the testing phase, our CO-GNN network is compared against five benchmarks.

%benchmark
The benchmarks we compare the CO-GNN networks with are as follows:

%There are some benchmarks we compare the neural network with:
\begin{itemize}

\item{\textit{Benchmark 1. GNN $\&$ Average Power:} To evaluate the effectiveness of joint optimization in CO-GNN, we conduct comparative GNN-based optimization experiments. Specifically, we design a neural network model closely resembling CO-GNN to maximize the sum secrecy rate. The model's configuration restricts optimization variables to beamforming and phase shifts, omitting the power allocation factor. We allocate power evenly between the two users, resulting in a uniform power allocation factor of $\left[0.5, 0.5\right]$.}
\item{\textit{Benchmark 2. GNN $\&$ Random IRS:} Given the random IRS, the GNN network is trained to optimize both the power allocation factor and beamforming to maximize the sum secrecy rate. }
\item{\textit{Benchmark 3. GNN $\&$ Omni-beam:} Given the omnidirectional beam, the GNN network is trained to optimize the power allocation factor and phase shifts to maximize the sum secrecy rate. }
\item{\textit{Benchmark 4. Alternating Optimization:} In the alternating optimization approach proposed in \cite{8982186}, we conduct joint optimization in three phases to enhance the secrecy rate. Initially, beamforming is optimized using the weighted minimum mean-squared error (WMMSE) algorithm \cite{5756489}. Next, phase optimization is performed employing the Riemannian conjugate gradient (RCG) algorithm \cite{10.5555/2627435.2638581}. Subsequently, the problem simplifies, enabling the application of a genetic algorithm for optimizing the power allocation factors.}
% \item{\textit{Benchmark 5. Reinforcement Learning:} The RL methodology proposed in \cite{9880822} was adopted and customized for the considered communication scenario to enable simultaneous optimization of beamforming vectors, IRS phase shift matrices, and power allocation parameters, while establishing the sum secrecy rate as the optimization objective. }
% \item{\textit{Benchmark 5. Hardware impairments:} Communication nodes are vulnerable to hardware impairments that can significantly degrade network performance. Addressing and mitigating the effects of these impairments are crucial for developing resilient communication systems. Therefore, this study assesses the ability of our CO-GNN networks to generalize to hardware impairment data and evaluates the system's robustness in terms of security performance. The signals and optimization objective, affected by hardware impairments, are described as (\ref{eq: impairments_1}) and (\ref{eq:goal_hi}). }

\end{itemize}

\subsection{Results and Analysis}

%To investigate the convergence of the adopted CO-GNN algorithm, we perform several experiments with different critic learning rates and batch sizes, where the results are illustrated in Fig. \ref{fig:loss_epoch}. We focus on the impacts of the critic learning rates since the performance of the CO-GNN algorithms is strongly determined by the approximation accuracy of the action-value function. From Fig. \ref{fig:loss_epoch}, it can be seen that the learning rate $1 \times 10^{-4}$ has lower loss, that is, higher secrecy rate than the learning rate $1 \times 10^{-5}$. Moreover, though batch size $32$ needs more epochs to reach the convergence, it always get the lowest loss with given learning rate. Hence, we adopt a batch size of $32$ and a critic learning rate of $1 \times 10^{-4}$ in the subsequent experiments to get the best optimization result.

\begin{figure} % Single column figure
	\includegraphics[width=\linewidth]{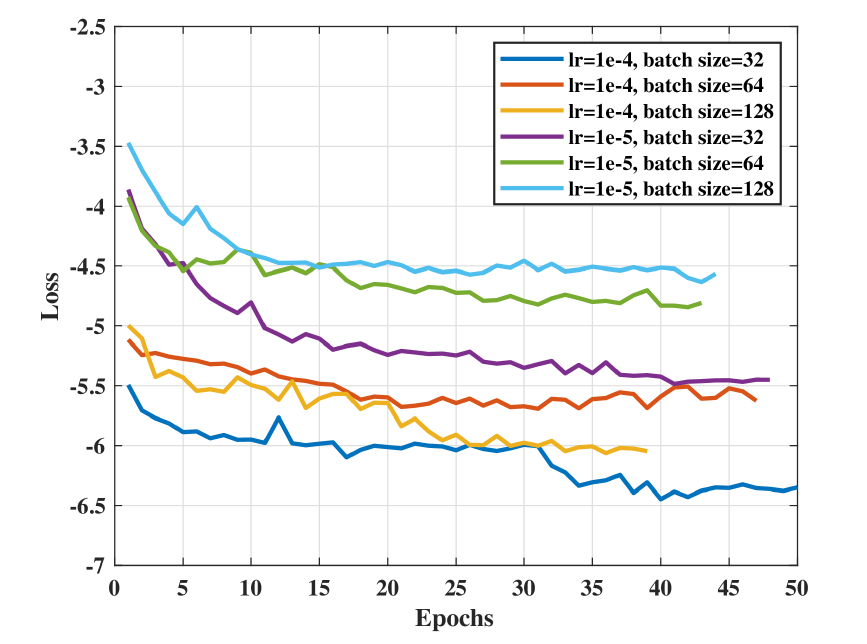}
    \caption{The convergence of the proposed CO-GNN networks}
	\label{fig:loss_epoch}
\end{figure}

To assess the convergence of the CO-GNN algorithm, we conduct several experiments varying the critic learning rates and batch sizes in the external eve scenario, with results depicted in Fig. \ref{fig:loss_epoch}. Our focus lies on the impact of critic learning rates, as the performance of CO-GNN algorithms heavily relies on the accuracy of the action-value function approximation. As depicted in Fig. \ref{fig:loss_epoch}, a learning rate of $1 \times 10^{-4}$ yields lower loss, indicating a higher secrecy rate compared to the learning rate of $1 \times 10^{-5}$. Additionally, although a batch size of $32$ requires more epochs to converge, it consistently exhibits the lowest loss with the given learning rate. Consequently, we opt for a batch size of $32$ and a critic learning rate of $1 \times 10^{-4}$ in subsequent experiments to achieve optimal optimization results.

\begin{figure} % Single column figure
	\includegraphics[width=\linewidth]{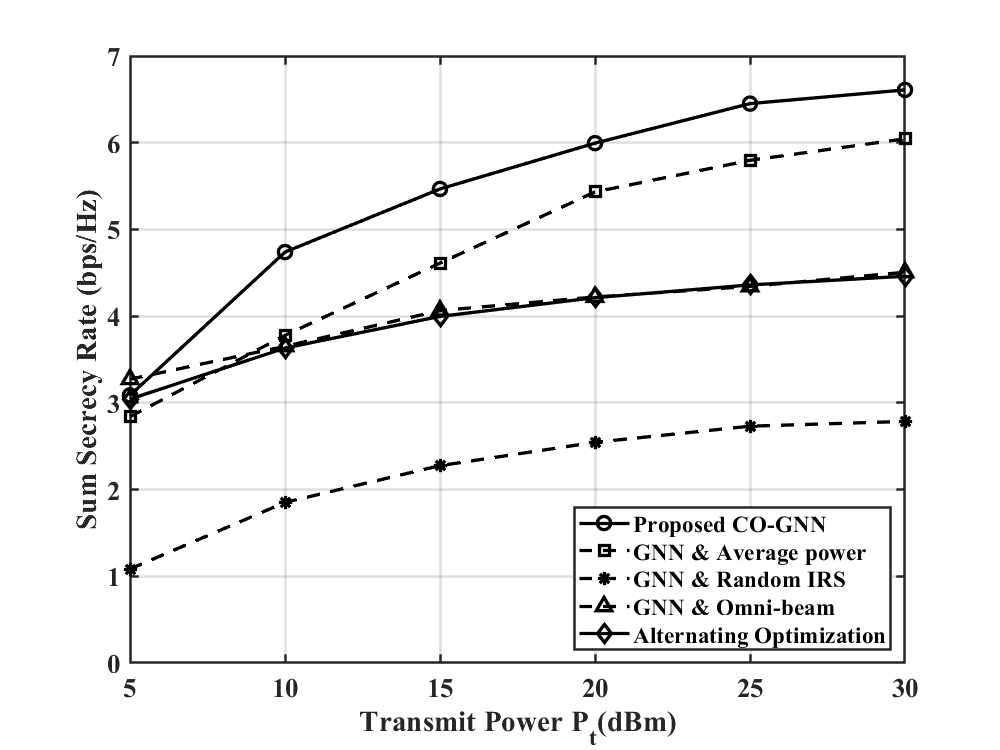}
    \caption{The sum secrecy rate versus transmit power $P_t$ under external eavesdropping scenario, with $M=5$, $N=100$, and $K=2$.}
	\label{fig:sr_power_external}
\end{figure}

Fig. \ref{fig:sr_power_external} illustrates the sum secrecy rate of various schemes concerning the transmit power $P_t$ when $M=5$, $N=100$, and $K=2$ in an external eavesdropper scenario. 
Simulation results demonstrate that the sum secrecy rate monotonically increases with the transmit power Pt. Remarkably, the proposed CO-GNN scheme achieves the highest secrecy rate among all compared methods, significantly outperforming conventional benchmarks. In scenarios with external eavesdroppers, the sum secure rate of CO-GNN exhibits approximately 40\% improvement over the AO scheme. This observation is consistent with conventional multiuser Multiple-Input Single-Output (MISO) systems. Through the joint optimization of transmit beamforming, power allocation, and phase shifts, the co-channel interference can be designed to damage the Eve, leading to performance enhancement with increasing $P_t$. Notably, the random IRS scheme exhibits the lowest rate. This underscores the significance of incorporating an IRS and optimizing its parameters to significantly enhance the system's secrecy rate. Moreover, both the 'GNN $\&$ Omni-beam' and the 'Alternating Optimization' scheme demonstrate nearly identical performance. This suggests that the similarity in optimization efforts of both algorithms results in their convergence towards the same stationary solution with high probability.

\begin{figure} % Single column figure
	\includegraphics[width=\linewidth]{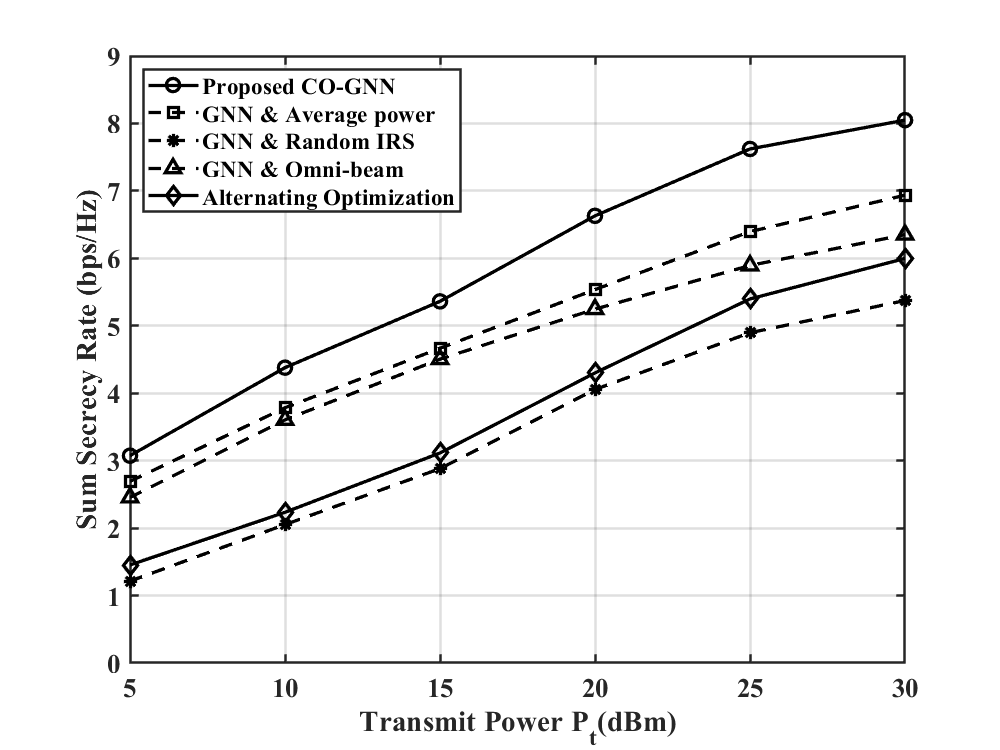}
    \caption{The sum secrecy rate versus transmit power $P_t$ under internal eavesdropping scenario, with $M=5$, $N=100$, and $K=2$.}
	\label{fig:sr_power_internal}
\end{figure}

Fig. \ref{fig:sr_power_internal} illustrates the sum secrecy rate as a function of power $P_t$ under an internal eavesdropping scenario with $M=5$, $N=100$, and $K=2$. In contrast to external eavesdropping, the secrecy rate exhibits a more pronounced increase in the internal eavesdropping scenario. We hypothesize that this is due to the node targeted for internal eavesdropping being treated as a legitimate node before in the model. Consequently, the neural network can effectively enhance its link performance. Conversely, when the node is acting as an eavesdropper, the neural network can more effectively suppress its activity, resulting in a higher level of security. Simulation results demonstrate that under internal eavesdropping scenarios, the CO-GNN scheme achieves approximately 30\% higher secrecy rate than the conventional AO approach. Additionally, the 'GNN $\&$ Random IRS' scheme, despite being the lowest-performing, demonstrates similar effects to the "Alternating Optimization" scheme, highlighting the advantages of employing GNN for optimization in the context of internal eavesdropping scenarios.

%Fig. \ref{fig:sr_power_internal} plots the sum secrecy rate versus power $P_t$ under internal eavesdropping scenario when $M=5$, $N=100$. Unlike external eavesdropping, the secrecy rate increases even more in the internal eavesdropping scenario. We speculate that because the node of internal eavesdropping turned out to be a legitimate node in the model. When it is a legitimate node, the neural network can effectively improve its link performance, but when it is an eavesdropper, the neural network can also suppress it more effectively, so it leads to a higher security rate. Moreover, the lowest-performing "GNN $\&$ Random IRS" scheme has similar effects to the "Alternating Optimization" scheme, which shows the advantages of using GNN for optimization under internal eavesdropping scenario.

% \begin{figure} % Single column figure
% 	\includegraphics[width=\linewidth]{fig/srate_power_internal_v5.eps}
%     \caption{The sum secrecy rate versus transmit power $P_t$ under internal eavesdropping scenario, with $M=5$, $N=100$, and $K=2$.}
% 	\label{fig:sr_power_internal}
% \end{figure}

\begin{figure} % Single column figure
	\includegraphics[width=\linewidth]{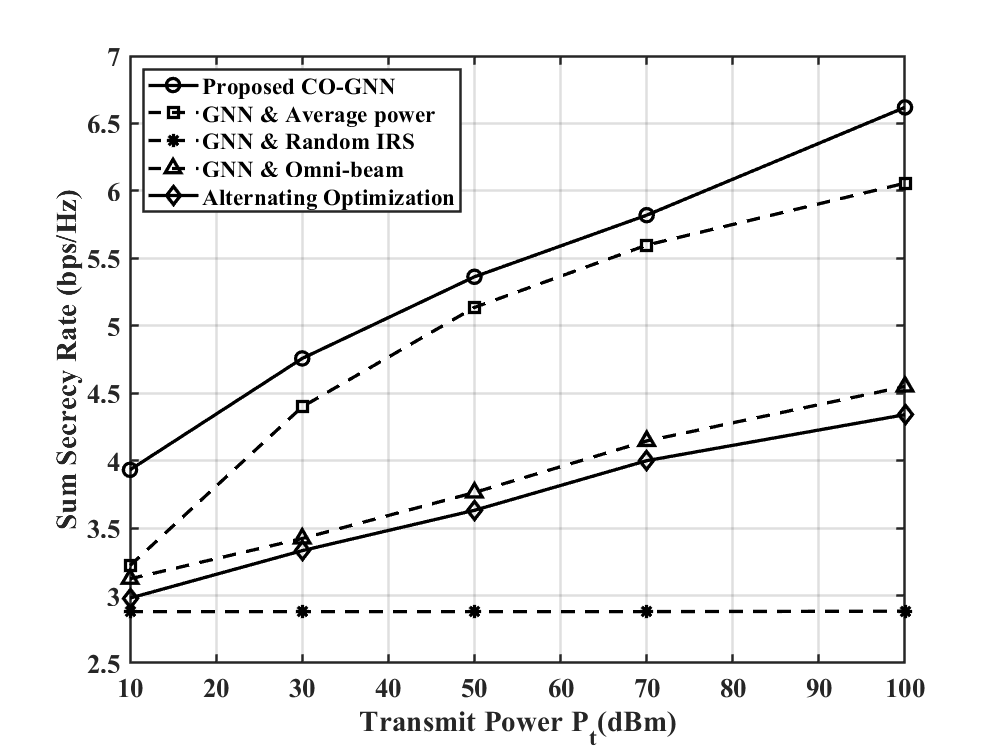}
    \caption{The sum secrecy rate versus the number of reflecting elements $N$ under external eavesdropping scenario, with $M=5$, $K=2$, and $P_t=30$dBm.}
	\label{fig:sr_N_external}
\end{figure}

Fig. \ref{fig:sr_N_external} presents a comparison of the sum secrecy rate against the number of elements in the IRS $N$ in an external eavesdropping scenario with $M=5$, $K=2$ and $P_t=30$dBm. It is noticeable that the rate for the 'GNN $\&$ Random IRS' scheme remains relatively constant, exhibiting insensitivity to variations in the number of IRS elements, whereas all IRS-assisted schemes achieve remarkable performance gains as $N$ increases. By deploying an IRS with a greater number of reflecting elements, we can effectively enhance the user links and suppress external eavesdropping, ensuring the security of the two legitimate users.
% Furthermore, the curve depicting hardware impairments closely resembles our CO-GNN curve, suggesting that our CO-GNN networks retain their resilience against hardware impairments even under the influence of $N$.

%Fig. \ref{fig:sr_N_external} compares the sum secrecy rate as a function of the number of elements in IRS $N$ under an external eavesdropping scenario with $M=5$, $P_t=30$. It can be observed that the rate for "GNN $\&$ Random IRS" nearly keeps constant which is insensitive to changes in the number of IRS, while all the IRS-aided schemes acheive remarkable performance gain as $N$ increases. By deploying the IRS with a larger number of reflecting elements, we can suppress the external eavesdropping and guarantee the security of two legitimate users effectively.

\begin{figure} % Single column figure
	\includegraphics[width=\linewidth]{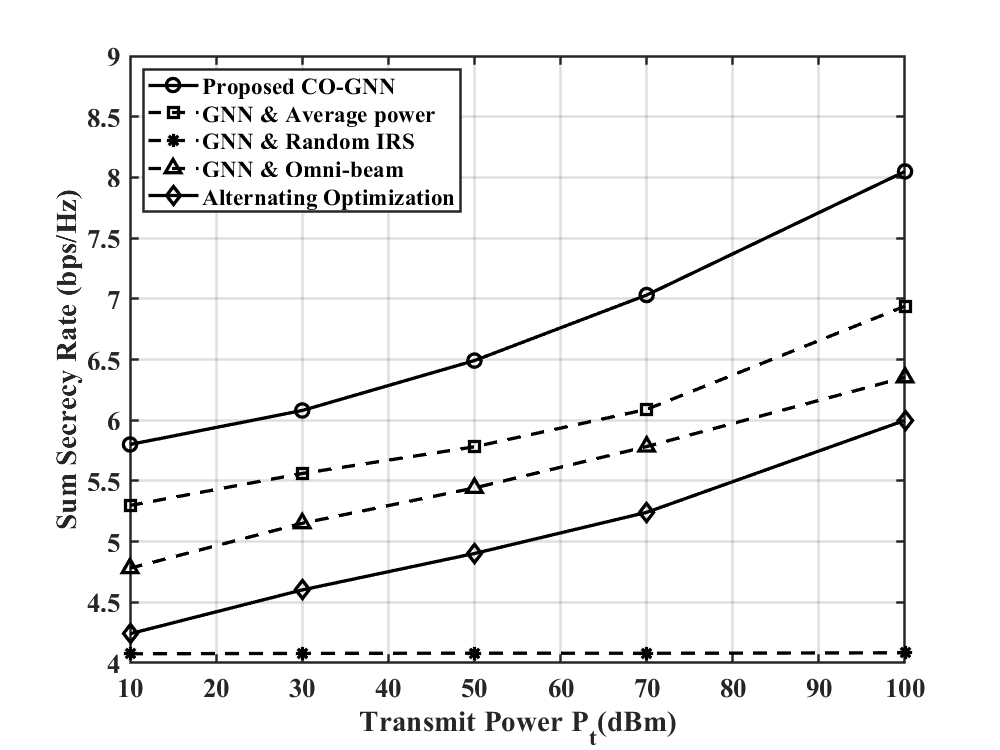}
    \caption{The sum secrecy rate versus the number of reflecting elements $N$ under internal eavesdropping scenario, with $M=5$, $K=2$, and $P_t=30$dBm.}
	\label{fig:sr_N_internal}
\end{figure}

Fig. \ref{fig:sr_N_internal} elucidates the variation of the sum secrecy rate concerning the number of elements in the IRS $N$ within an internal eavesdropping scenario with parameters $M=5$, $K=2$ and $P_t=30$dBm. With the exception of the relatively insensitive 'GNN $\&$ Random IRS' scheme, all GNN optimization schemes outperform the 'Alternating Optimization' scheme, with our proposed CO-GNN demonstrating the highest sum secrecy rate. Furthermore, in the context of internal eavesdropping, the trend observed in this figure closely mirrors the curve depicted in Fig. \ref{fig:sr_power_internal}, suggesting a similarity in the effects of power and the number of IRS elements on the sum secrecy rate under internal eavesdropping scenarios.

%Fig. \ref{fig:sr_N_internal} elaborate how the sum secrecy rate varies with the number of IRS elements $N$ under internal eavesdropping scenario with $M=5$, $P_t=30$. Apart from the nsensive "GNN $\&$ Random IRS" scheme, all the GNN optimization scheme outperforms the AO scheme, and our CO-GNN achieve the best sum secrecy rate. Moreover, in the case of internal eavesdropping, Fig. \ref{fig:sr_power_internal} is similar to the curve of this figure, and it can be inferred that the effect of power and the number of IRS is similar under internal eavesdropping scenario.

% \begin{figure} % Single column figure
% 	\includegraphics[width=\linewidth]{fig/srate_N_internal_v3.eps}
%     \caption{The sum secrecy rate versus the number of reflecting elements $N$ under internal eavesdropping scenario, with $M=5$, $K=2$, and $P_t=30$dBm.}
% 	\label{fig:sr_N_internal}
% \end{figure}
\begin{figure} % Single column figure
	\includegraphics[width=\linewidth]{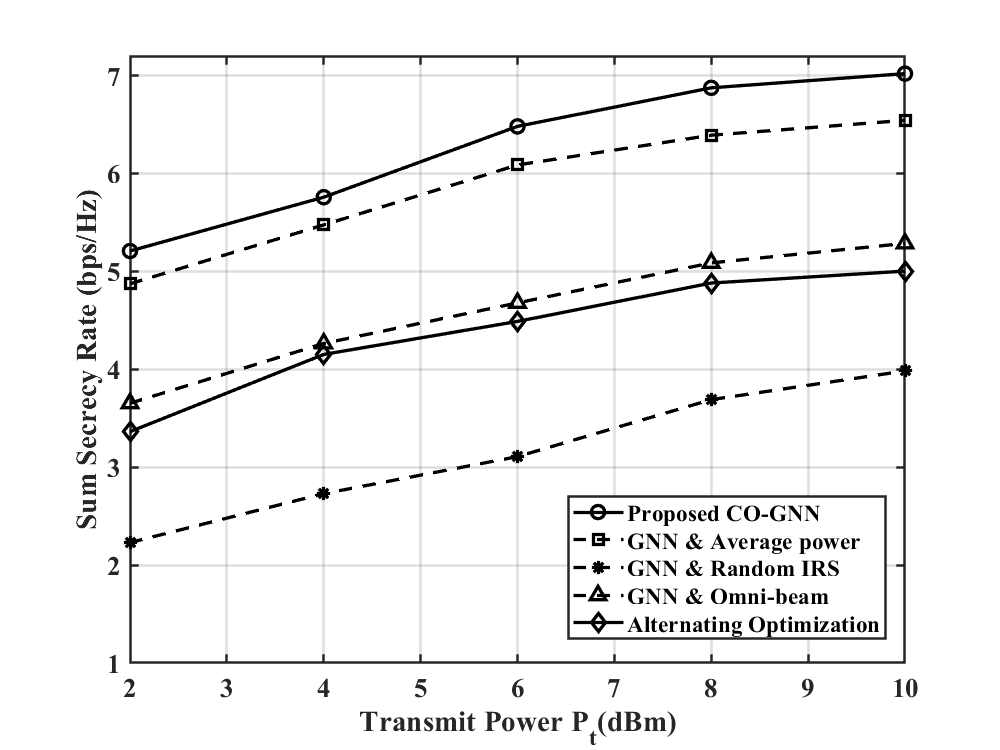}
    \caption{The sum secrecy rate versus the number of antennas $M$ under external eavesdropping scenario, with $K=2$, $N=100$ and $P_t=30$dBm.}
	\label{fig:sr_M_external}
\end{figure}

\begin{figure} % Single column figure
	\includegraphics[width=\linewidth]{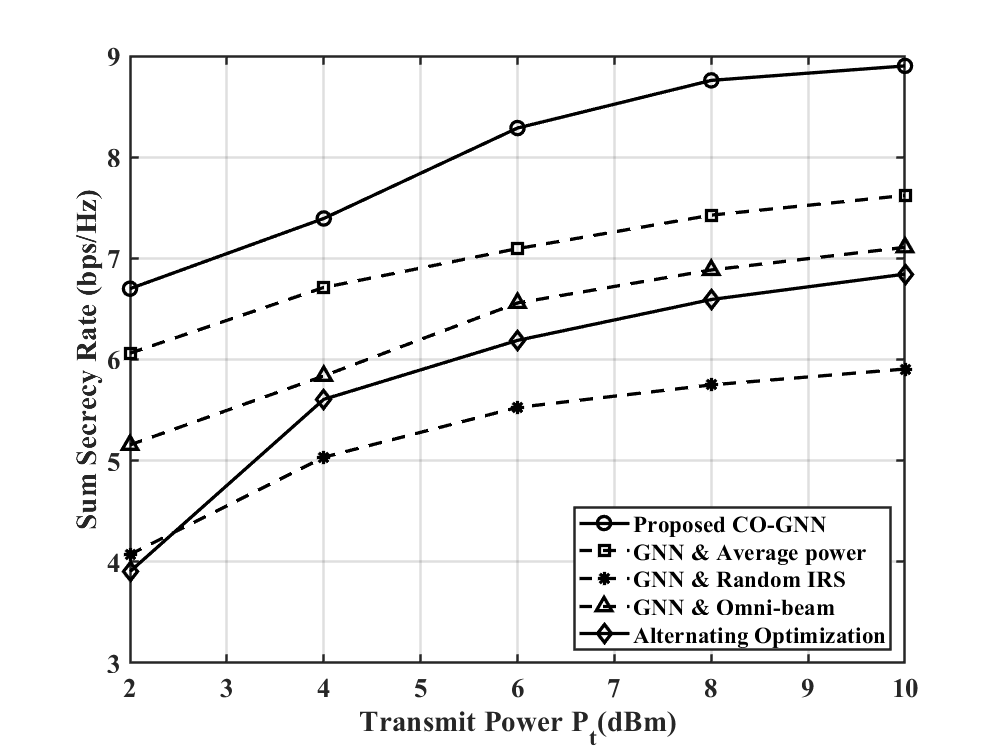}
    \caption{The sum secrecy rate versus the number of antennas $M$ under internal eavesdropping scenario, with $K=2$, $N=100$ and $P_t=30$dBm.}
	\label{fig:sr_M_internal}
\end{figure}

Fig. \ref{fig:sr_M_external} and Fig. \ref{fig:sr_M_internal} delineate the influence of the number of transmit antennas $M$ on the sum secrecy rate under the conditions of $K=2$, $N=100$ and $P_t=30$dBm. These figures reveal a monotonic increase in the sum secrecy rate as the number of transmit antennas increases. It can be attributed to the ability of additional antennas to bolster the strength of the main links and provide a more centralized approach, thereby confounding potential eavesdroppers through optimized beamforming, power allocation, and phase shifts. It is noteworthy that even with a limited number of antennas, the sum secrecy rate remains high across all experimental scenarios. This resilience is attributed to the combined effect of the specified power level, $P_t=30$dBm, and the ample number of elements in the IRS, $N=100$, which collectively ensure the robust security of the system. 
% Meanwhile, our CO-GNN effectively mitigates the impact of hardware impairments.

%Fig. \ref{fig:sr_M_external} and Fig. \ref{fig:sr_M_internal} shows the impact of transmit antenna number $M$ on sum secrecy rate with $N=100$, $P_t=30$, which shows the sum secrecy rate monotonously increases as the number of transmit antennas. This is because more antennas can enhance the main links and more centralized to confuse the Eve with the presence of optimized beamforming, power allocator and phase shifts. It is important to note that even with a small number of antennas, the sum secrecy rate is still high in all experiments, because the power, $P_t=30$, and the number of IRS, $N=100$, can ensure the safety of the system well.   

\subsection{Interpretation of Optimization of CO-GNN}
%为了验证我们的图神经网络的联合优化的有效性，我们将被优化的IRS进行了可视化，以证明波束赋形和IRS被优化到了理想的方向和角度。对该结果的探索离不开各节点的角度设置，所以我们令xx为到达角和仰角xx.
%To validate the efficacy of the joint optimization in our CO-GNN, we present visualizations of the optimized results to showcase the optimized directions and angles of beamforming and phase shifts. The analysis of these outcomes intricately hinges on the angular configurations of individual nodes. We mainly focus on the indirect link, since the links from users to IRS and from IRS to BS includes the manifestation of both beamforming and phase shifts. Therefore, we assign $\phi_1$, $\phi_2$ as the AoD from the IRS to the BS and the AoA from the users to the IRS, respectively. Correspondingly, the elevation angles are designated as $\theta_1$ and $\theta_2$.

% \begin{figure} % Single column figure
% 	\includegraphics[width=0.85\linewidth]{pic/visual_3.png}
%     \caption{The visual simulation setup of experimental configuration}
% 	\label{pic:visual}
% \end{figure}

\begin{figure*}[b] % Two column figure (notice the starred environment)
	\centering
    \subfloat[$N=10$.]{\includegraphics[scale=0.4]{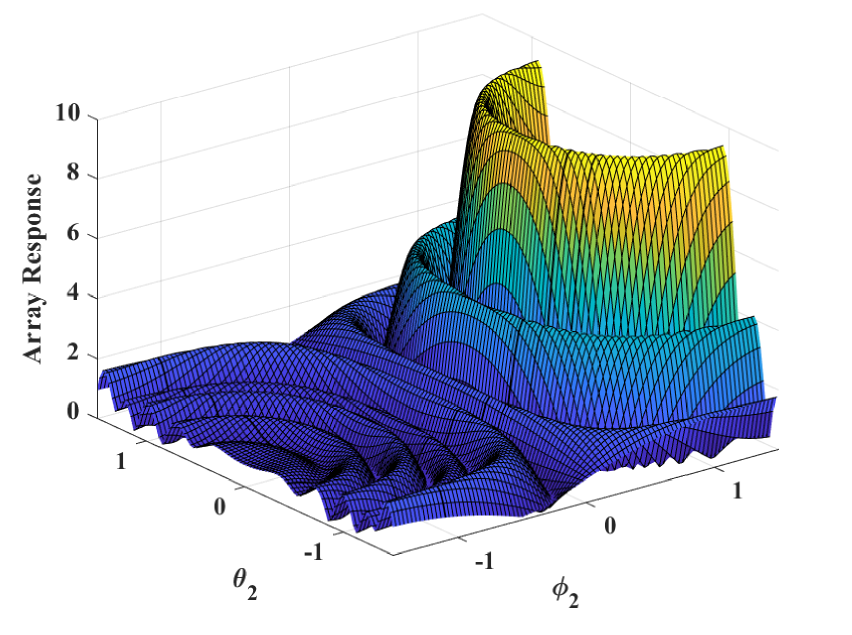}}
    \subfloat[$N=50$.]{\includegraphics[scale=0.4]{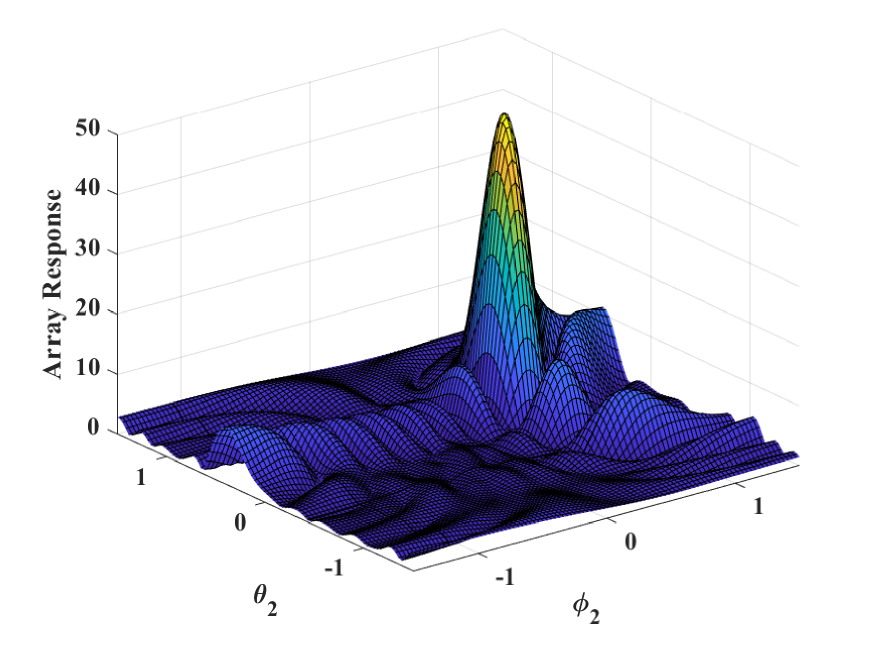}}
    \subfloat[$N=100$.]{\includegraphics[scale=0.4]{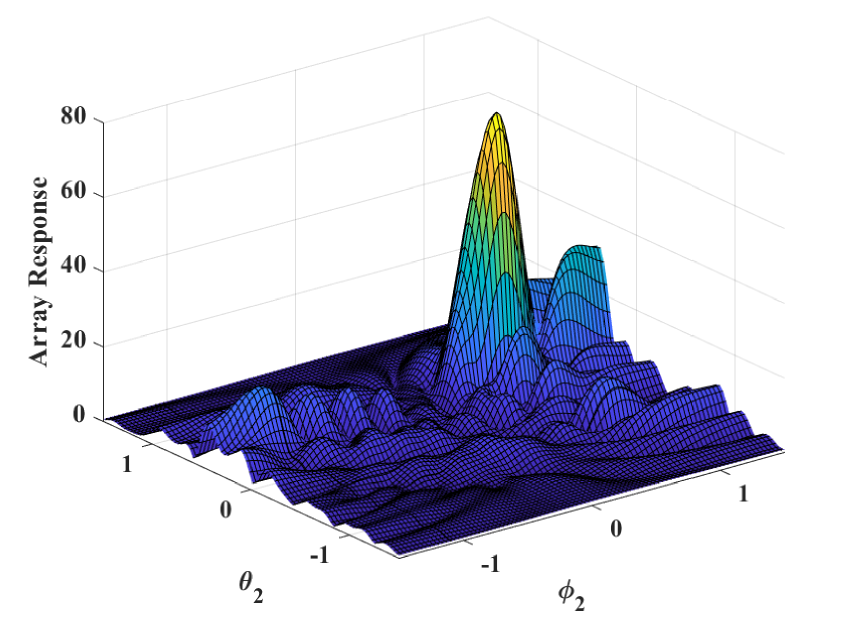}}
 
    \caption{Array response of IRS obtained from CO-GNN over two users with $M=5$, $3$-D view. The optimal $(\phi_2,\eta_2)=(0.4636,0.4082), (0.6435,0.3714)$. }
    \label{fig:visual_2}
\end{figure*}

To substantiate the effectiveness of joint optimization within our proposed CO-GNN framework, we present visualizations demonstrating the optimized outcomes. These visualizations aptly illustrate the refined directions and angles relevant to beamforming and phase shifts. The detailed analysis of these results heavily relies on the angular configurations of each node. Accordingly, we define $\phi_1$, $\phi_2$ as the AoD from the IRS to the BS and AoA from the users to the IRS, respectively. The corresponding elevation angles are denoted as $\eta_1$ and $\eta_2$.

From the steering vectors (\ref{eq:a_IRS}), (\ref{eq:GLos}), (\ref{eq:sincos_phi1}) and (\ref{eq:sin_1}), we can get the array response of IRS as follows:

\begin{equation}
    {\rm A}\left( {{\phi _1},{\eta _1},{\phi _2},{\eta _2}} \right) = \left| {{\mathbf{\Phi}^H}{e^{j\pi \left( {{i_1}\left( n \right){\tau _1} + {i_2}\left( n \right){\tau _2}} \right)}}} \right|,
\end{equation}
where ${\tau _1} = \sin \left( {{\phi _2}} \right)\cos \left( {{\eta _2}} \right) - \sin \left( {{\phi _1}} \right)\cos \left( {{\eta _1}} \right)$ and ${\tau _2} = \sin \left( {{\eta _2}} \right) - \sin \left( {{\eta _1}} \right)$.

%oa_irs_y_k aoa_irs_z_k [0.408248290463863, 0.5570860145311556] [-0.408248290463863, -0.3713906763541037]phi (0.982793723247329, -0.9827937232473292, array([0.46364761, 0.64350111]))
%所以我们可以得到不同IRS数量下，角度对阵列响应的影响。
%So we can get the effect of IRS angles on the array response at different IRS numbers. The location of BS and IRS are set as $(0,0,0)$ and $(20,30,0)$. And the two users are located at $(40,40,-10)$ and $(40,45,-10)$. Then the $(\phi_2,\theta_2)=(0.46364,0.40824), (0.6435,0.37139)$. We load the trained model with optimized beamforming, power allocator and phase shifts to get the array response of IRS. The results are depicted in Figxx, where xx are on the Top-down view and xx are the $3$-D view with $N=10,50,100$. From these figures, we can see that the peaks matches the angles $\phi_2$ and $\theta_2$ corresponding to the two users. Meanwhile, as the number of IRS increases, the poeaks are higher and more focus, meaning that more IRS elements have higher control over the beamforming, resulting in better optimization and sum secrecy rate. Thus, our CO-GNN indeed learns to make the joint optimization maximize the sum secrecy rate across the two users. 

\begin{figure}[h] % Single column figure
	\includegraphics[width=0.85\linewidth]{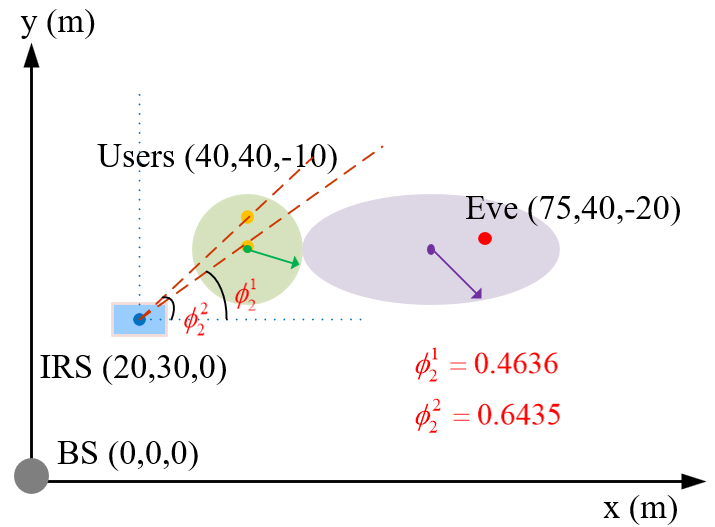}
    \caption{The visual simulation setup of experimental configuration}
	\label{pic:visual}
\end{figure}

To investigate the impact of angles on the array response at varying IRS numbers, we fixed the location of the BS at $(0,0,0)$ and the IRS at $(20,30,0)$. Furthermore, we positioned two users at $(40,40,\--{10})$ and $(40,45,\--{10})$. The visualization of this experimental configuration is shown in Fig. \ref{pic:visual}. We assigned $(\phi_2,\eta_2)$ values of $(0.4636,0.4082)$ and $(0.6435,0.3714)$ to these users. Utilizing our trained model with optimized beamforming, power allocation, and phase shifts, we obtain the array response of the IRS. The results are presented in Fig. \ref{fig:visual_2}, which comprises a 3D view for $N=10,50,100$. From these figures, it is evident that the peak values precisely align with the angles $\phi_2$ and $\eta_2$, corresponding to the two users. Notably, as the number of IRS elements increases, the peaks become more pronounced and concentrated to the angles of both legitimate users. This signifies that a greater number of IRS elements exert higher control over the beamforming, enabling a more focused and directed transmission of signals towards the users, while effectively mitigating the signal leakage towards the eavesdropper. Hence, our proposed CO-GNN framework effectively integrates joint optimization techniques, leading to maximized sum secrecy rates for both users.

%\begin{figure} % Single column figure
%	\includegraphics[width=\linewidth]{fig/impairments.eps}
%    \caption{hardwareimpairments}
%	\label{fig:sr_power_impairments}
%\end{figure}

%When the P value is between 0.01 and 0.05, it indicates a significant difference. In this case, the P-value, $0.029723<0.05$, means the hardware impairments data is different from the original input data. 

\begin{comment}
    \begin{table}[!ht]
    \centering
    \begin{tabular}{c|c|c|c|c}
    \hline
        Data & Number of observations & F & P-value & F crit \\ \hline
        y\_origin & num\_samples*1000 & 4.79467 & 0.029723 & 3.88934 \\ %\hline
        y\_impairments & num\_samples*1000 & ~ & ~ & ~ \\ \hline
    \end{tabular}
\end{table}
\end{comment}

\subsection{Computational Complexity Analysis}
This subsection investigates the computational complexity of different schemes. For the GNN-based schemes, since there is no strict time limit at the offline training stage, we only consider the computational complexity at the online test stage in the CPU. The computational complexity analysis of different schemes are as follows.

\begin{itemize}
    \item{For our CO-GNN scheme, the computational complexity is ${\rm O}\left( {\left( {3 \times 3 + L} \right){d_{MLP}}{T_R}} \right)$, where $ {3 \times 3}$ means the number of input and output layers of three optimized parameters, $L=8$ is the number of network layers, $d_{MLP}=512$ is the dimension of linear embedding and $T_R$ is the time of calculate sum secrecy rate. }
    \item{The computational complexity of \textit{GNN $\&$ Average Power}, \textit{GNN $\&$ Random IRS}, \textit{GNN $\&$ Omni-beam} is ${\rm O}\left( {\left( {3 \times 2 + L} \right){d_{MLP}}T_R^{''}} \right)$ since the optimization goal is two in these schemes. Meanwhile, there are differences in the $T_R^{''}$ in these schemes, leading to slight variations in computational complexity.  }
    \item{As for the AO algorithm, the complexity of the WMMSE algorithm is ${\rm O}\left( {{I_\lambda }{I_w}K{M^3}} \right)$, where $I_\lambda$ and $I_w$ are the iteration numbers of searching $\lambda$ and the three-step updating loop. Meanwhile, the complexity of the RCG algorithm is decided by the Euclidean gradient, which is computed as ${\rm O}\left( {{K^2}{N^2}} \right)$. The complexity of genetic algorithm is depends on the Population size $P_o$ and evolutionary generations $t$. So the total complexity of the alternating optimization is ${\rm O}\left( {{I_\lambda }{I_w}K{M^3} + {K^2}{N^2} + {P_o}t} \right)$.}
\end{itemize}

\begin{figure} 
	\includegraphics[width=\linewidth]{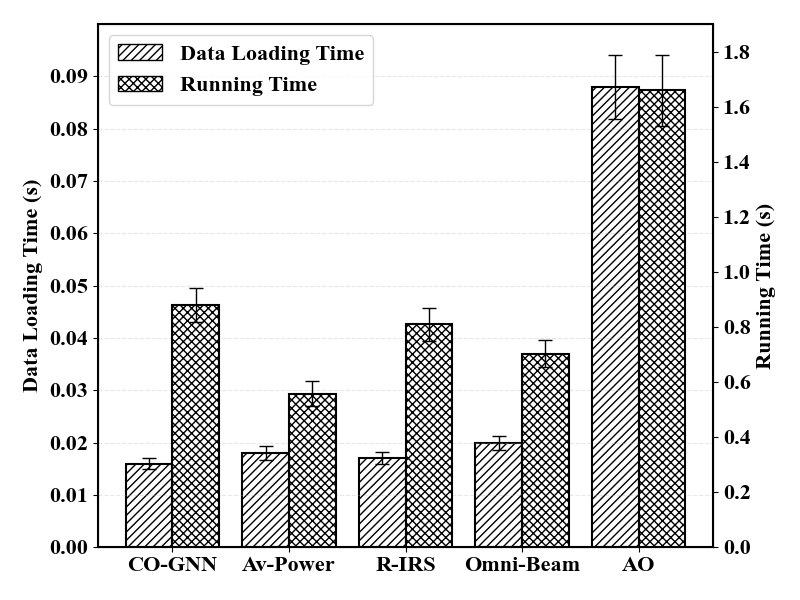}
    \caption{Computing time of different schemes.}
	\label{fig:compute_time}
\end{figure}

To intuitively observe the computational complexity across different schemes, Fig. \ref{fig:compute_time} shows the data loading time and the running time of different schemes in the CPU. Notably, the figure reveals that our CO-GNN approach significantly outperforms the AO scheme in both metrics. In terms of data loading time, which encapsulates the duration for importing test data and pre-trained models, the AO algorithm's slightly prolonged model import time contributes to its marginally higher overall loading time compared to other schemes. As for running time, the AO scheme necessitates separate optimization of three parameters, which results in a higher execution time compared to the neural network model, which directly derives the optimized parameters.

%To intuitively observe the computational complexity of different schemes, Fig. \ref{fig:compute_time} shows the data loading time and the running time of different schemes in the CPU. It can be observed that the time of testing data loading and running time of our CO-GNN is significantly lower than that of the AO scheme. Data load time includes the time it takes to import test data as well as trained models. Because the model import time of the AO algorithm is slightly longer, the import time is slightly longer than that of other schemes. For the running time, since AO needs to optimize the three parameters separately, its running time is higher than that of the neural network model that directly obtains the optimized parameters.

\section{Conclusion}

In this paper, we have developed a secure model for an IRS-assisted NOMA transmission system, accounting for both external and internal eavesdroppers in heterogeneous networks, as well as potential hardware impairments. Furthermore, we introduce a novel joint optimization approach, CO-GNN, which is designed to optimize the heterogeneous resources and obtain the optimized transmit beamforming at BS, power allocation for NOMA users, and phase shifts of IRS, aiming to maximize the sum secrecy rate of legitimate users. Our CO-GNN exhibits a standardized formulation and low implementation complexity, eliminating the need for explicit mathematical representations of wireless systems and intricate channel estimation procedures. This simplifies the scaling process to accommodate diverse system configurations.

Simulation results underscore the necessity and efficacy of jointly optimizing beamforming, power allocation, and phase shifts. Additionally, as transmit power, the number of IRS reflecting elements, or transmit antennas increase, the security of our model experiences a significant enhancement. Notably, compared with conventional AO schemes, the CO-GNN achieves superior security performance while maintaining low computational complexity. These findings not only establish the superiority of our proposed method but also highlight its practicality and scalability in real-world scenarios.

Future research will extend the current framework by: rigorously validating the theoretical security boundaries of CO-GNN against heterogeneous attack strategies in complex threat scenarios to establish a generalized anti-jamming model; designing real-time sensing-based adaptive IRS reconfiguration algorithms to address practical deployment challenges such as IRS occlusion, mobility, and environmental perturbations, thereby enhancing system resilience under non-ideal channel conditions; and exploring deep integration mechanisms between CO-GNN and physical-layer security techniques, such as artificial noise injection and cryptographic encoding, to construct a cross-layer cooperative optimization framework that maximizes eavesdropper obfuscation through dynamic resource allocation while ensuring quality-of-service for legitimate users. These efforts will systematically bridge the gap between theoretical models and engineering practices, facilitating reliable deployment of secure IRS-NOMA networks in complex heterogeneous environments.

\bibliographystyle{IEEEtran}
\bibliography{reference}{}

\vfill

\end{CJK}

\end{document}